\documentclass{aa}  

\usepackage{graphicx}
\usepackage{txfonts}
\usepackage{amsmath,amsfonts,amssymb}
\usepackage{textcomp,gensymb}
\usepackage[T1]{fontenc}
\usepackage{aecompl}
\usepackage{ulem}
\usepackage[dvipsnames,svgnames]{xcolor}
\usepackage{color}
\usepackage{times}
\usepackage{subfig}
\usepackage{stfloats}
\usepackage[utf8]{inputenc}
\usepackage{hyperref}
\usepackage{multirow}
\usepackage{natbib}
\usepackage{placeins}
\usepackage{booktabs}
\usepackage{orcidlink} 
\usepackage{soul}

\begin{document} 
    \title{Protoplanetary Disc Population Synthesis}
    \subtitle{I. Linking mass-dependent viscosity, star formation history and environment to observations}
   
    \authorrunning{J. L. Gomez et al.}
    \author{Jose L. Gomez \orcidlink{0009-0009-4329-2916} \inst{1,2},      
          Octavio M. Guilera \orcidlink{0000-0001-8577-9532}\inst{1,2},
          Marcelo M. Miller Bertolami \orcidlink{0000-0001-8031-1957} \inst{1,2},
          Elisa Castro Martínez \orcidlink{0009-0001-1312-9503} \inst{3} \and \\
          Mar\'{\i}a Paula Ronco \orcidlink{0000-0003-1385-0373} \inst{1,2}
          }
   \offprints{Jose L. Gomez}    
   \institute{Instituto de Astrof\'{\i}sica de La Plata, CCT La Plata, CONICET-UNLP, Paseo del Bosque S/N, B1900FWA La Plata, Argentina 
    \and
    Facultad de Ciencias Astronómicas y Geofísicas, UNLP, Paseo del Bosque S/N, B1900FWA La Plata, Argentina 
    \and 
    Univ. Grenoble Alpes, CNRS, IPAG, 38000 Grenoble, France \\
    \email{jlgomez@fcaglp.unlp.edu.ar}
    }
   
   \date{Received XX, 2025; accepted XX, 2025}

  \abstract
  {Protoplanetary discs are the birthplaces of planets. Recent studies highlight the importance of stellar mass sampling when determining disc lifetimes through the observed fraction of stars with discs. Low-mass stars tend to host discs with average lifetimes exceeding 5 Myr, providing sufficient time for planet formation via solid accretion. In addition, observations show a strong correlation between stellar (and substellar) mass and accretion rate, typically following the relation $\dot{M} \propto M_\star^2$.}   
  {This study aims to identify the optimal parameters of a protoplanetary disc evolution model capable of reproducing both the observed disc fractions and the mass accretion rates in young stellar populations.}
  {We conducted a population synthesis study where disc evolution was modelled exploring different values of the viscosity parameter $\alpha$ as a function of stellar mass. The evolution of the discs was modelled through viscous accretion and photoevaporation (internal and external). Initial disc masses and radii were drawn from observationally motivated statistical distributions. We also introduce in our population synthesis a stellar mass distribution and a star formation rate (SFR).}
  {Matching the observed disc fractions and accretion trends requires that the $\alpha$-viscosity parameter increases with stellar mass. External photoevaporation, as well as a systematic underestimation of the gas disc masses due to dust evolution and hidden solids, are required to reproduce the $\dot{M}-M_{\rm d}$ plane. Accounting for a time-dependent SFR enhances accretion in young clusters and extends disc lifetimes in older ones. In addition, introducing a stellar mass cut-off reproduces the distance-dependent biases seen in cluster disc fractions.}  
  {Our results indicate that both stellar and environmental dependencies are essential to explain the observed properties of protoplanetary discs. A stellar-mass–dependent viscosity is required to recover the $\dot{M}$–$M_\star$ relation, while external photoevaporation and extended star formation histories shape the distribution of accretion rates across different environments. These findings highlight the need for multi-faceted population synthesis models to connect disc evolution with planet formation pathways.}
    
  \keywords{Protoplanetary discs -- accretion discs -- methods: numerical}
   \maketitle

\section{Introduction}

Protoplanetary discs are a natural outcome of star formation and provide the material reservoir from which planetary systems form. Over the past two decades, major observational campaigns have dramatically expanded the available datasets of disc properties in nearby star-forming regions, enabling statistically robust studies of disc evolution across a wide range of stellar masses and environments \citep[e.g.][]{2016A&AManara, 2017A&ATazzari, manara2023ASPC}. These surveys have revealed a number of key empirical trends, including the fraction of stars with discs in young stellar clusters of different ages, and correlations between disc masses, stellar masses, and mass accretion rates, as well as a substantial intrinsic scatter whose physical origin is still debated.

Throughout their lifetimes, protoplanetary discs undergo significant evolution driven by viscous accretion, photoevaporation, and interactions with forming planets \cite[e.g.][]{Dullemond2006, Alexander2006,armitage_2009_evolution}. Disc lifetimes are therefore a fundamental parameter for planet formation, setting the timescale available for the growth of solids and the assembly of planetary cores \citep[e.g.][]{Venturini20Review, drazkowska2022planet}. Early studies of protoplanetary disc lifetimes suggested that discs typically disperse within 1 to 10 Myr, with an average lifetime of approximately 3 Myr \citep[e.g.][]{Haisch2001, MamajekE, Pfalzner_2014}. However, more recent analyses indicate that disc lifetimes may be significantly longer. \citet{Pfalzner_2022} highlight that observational samples of clusters are often biased by magnitude limits, resulting in an over-representation of high-mass stars in disc fraction estimates. Since protoplanetary discs around intermediate and massive stars evolve and dissipate faster \citep[e.g.][]{ribas2015protoplanetary, Kunitomo2021, Ronco2024}, this bias leads to a systematic underestimation of disc lifetimes. When accounting for stellar mass and distance biases, disc lifetimes appear to span a broader range, from 1 to 20 Myr, with most discs around low-mass stars persisting beyond 5 Myr \citep{Pfalzner_2022, 2024ApJPfalzner}.

The long-term evolution of protoplanetary discs is commonly described within two main theoretical frameworks: viscous accretion discs, parameterized through an effective viscosity \citep[e.g.][]{BellPringle1974,pringle1981accretion}, and magnetohydrodynamic wind-driven discs, in which angular momentum is primarily removed by large-scale magnetic torques \citep[e.g.][]{blandford1982,Suzuki2009}. Despite significant progress, the dominant transport mechanism operating in discs remains uncertain. Observational constraints on turbulence levels and accretion efficiencies suggest a wide range of possible disc viscosities \citep[e.g.][]{Ansdell017, Flaherty_2018, 2020A&A.A.R, Villenave2020, Villenave2022}, highlighting the need for statistical approaches to discriminate between competing models \citep[see][and reference therein, for a recent review]{2023Rosotti}.

Because individual discs cannot be observed throughout their entire evolutionary history, testing disc evolution models requires population-level approaches. Disc population synthesis has therefore emerged as a powerful framework to connect theoretical disc evolution models with observed disc demographics. \citet{Lodato2017MNRAS} introduced the concept of {\it disc isochrones} and demonstrated that the observed correlation between disc mass and stellar accretion rate arises naturally within the framework of viscous evolution when combined with a distribution of initial disc properties. This study laid the foundation for a growing number of population synthesis models aimed at interpreting the diversity of observed disc properties.

Subsequent works have expanded disc population synthesis to include additional physical processes and environmental effects. Internal photoevaporation has been explored as a mechanism to explain the observed spread in accretion rates and disc dispersal timescales \citep[e.g.][]{owen2012theoryfoto, Picogna2019, Somigliana2020, Somigliana2022MNRAS, Somigliana2024}, while the impact of external photoevaporation has been investigated in irradiated environments, where it can significantly affect disc masses and sizes \citep[e.g.][]{coleman2022dispersal, Coleman2024, Weder2025}. These works also highlight the importance of including both internal and external photoevaporative winds to accurately describe disc evolution. In environments with high ultraviolet radiation fields, transition discs are predicted to be rare \citep[e.g.,][]{Winter2018}, a trend that is consistent with their observed spatial distribution in the Orion Nebula Cluster \citep[e.g.,][]{Eisner2018}. The role of disc substructures in shaping disc observables has been studied by \citet{Toci2021, Delussu2024}, whereas the effects of infall, fragmentation, and disc instabilities have been explored, for example, by \citet{Schib2021, Schib2025}. Multiplicity has also been proposed as a key contributor to the observed scatter in accretion rates and disc masses \citep{Zagaria2022}.

Recent efforts have focused on constraining the initial conditions adopted in disc population synthesis models. In particular, \citet{emsenhuber2023towards} investigated how different assumptions on initial disc masses and radii affect the ability of population synthesis models to reproduce disc properties in nearby low-mass star-forming regions. They found that models including both internal and external photoevaporation often result in disc lifetimes shorter than observed, motivating further exploration of alternative initial conditions, environmental effects, and disc evolution pathways. In addition, \citet{coleman2022dispersal} demonstrated that ongoing star formation is necessary to match the observed disc fractions as a function of cluster age.

In this work, we build upon these previous studies and aim to reproduce the main observed properties of protoplanetary discs through a disc population synthesis framework in which discs evolve under viscous accretion and both internal and external photoevaporation. We focus on the gaseous component of discs and adopt initial conditions motivated by observations of Class 0/I systems. Particular attention is given to the role of observational biases, time-dependent star formation rates (SFR), and external irradiation environments. By systematically exploring these effects, we aim to identify the disc parameters and evolutionary pathways that best reproduce the observed disc populations across different star-forming regions.

The paper is structured as follows: in Sec.~\ref{sec:populsynt} we present the method adopted to perform the population synthesis and we describe the model used to compute the disc evolution. In Sec.~\ref{sec:constrains} we present the observational constraints for our population synthesis results. We then present the results of our study in Sec.~\ref{sec:result}, focusing on finding the best set of initial condition for the population, exploring how the $\alpha$ correlation with stellar masses, and the SFR affect the lifetime and mass accretion rates. This is followed by a discussion of our results in Sec.~\ref{sec:discussion}. Finally, we present our conclusions in Sec.~\ref{sec:conclusions}.

\section{Population synthesis model}
\label{sec:populsynt}

The modelling of the whole population of protoplanetary discs in a given young stellar population requires many different physical ingredients. First of all we need to know the distribution of stellar masses resulting from the fragmentation of the original molecular cloud. Then, we need to know how the masses and radii of the discs correlate with the mass of the central stars. Moreover, in order to predict the properties of protoplanetary discs at a given time we need to know the age of each individual disc, and how they change with time. Some of these properties are similar and well understood among several disc populations, and can be statistically derived from observations. Conversely, other parameters of the model will depend both on the specific details of a given stellar population and/or will depend on the underlying physical model of the disc, and need to be calibrated to reproduce specific populations. Below we describe how different ingredients have been adopted in our current population synthesis model. For each choice of the physical parameters of each population, 10000 random star-disc systems were simulated to derive the underlying statistical distributions.

\subsection{Initial stellar mass function} 

The initial stellar mass function implemented in this work is the one derived by \cite{kroupa2001}. The probability distribution of stellar masses ($\rm{M}_{\star}$) in each stellar population is given by:
\begin{equation}
    \xi (\rm{M}_{\star}) \propto
   \begin{cases} 
      M^{-0.3}_{\star},  & 0.01{~\rm M}_{\odot}\leq {M}_{\star}< 0.08{~\rm M}_{\odot},   \\
      M^{-1.3}_{\star},  & 0.08{~\rm M}_{\odot}\leq {M}_{\star}< 0.50{~\rm M}_{\odot},   \\
      M^{-2.3}_{\star},  & 0.5{~\rm M}_{\odot}\leq {M}_{\star}< 10.0{~\rm M}_{\odot},
   \end{cases}
\end{equation}
where stellar masses in our simulations are taken in the range 0.04-1.4 $M_{\odot}$. We take the normalization constant in each mass range so that the resulting probability distribution is continuous.

\subsection{Star formation rate} 

In most population synthesis models, it is often assumed that all stars in a cluster form simultaneously. However, observations suggest that star formation is an extended process, leading to an intrinsic age spread within stellar clusters. To account for this star formation history, and following the work of \cite{coleman2022dispersal}, we adopt a time-dependent SFR that initially remains constant for a million years ($\rm{SFR}_{0}$) and subsequently decays exponentially with a characteristic timescale $t_{\rm d}$. The SFR is given by
\begin{align}
    \rm{SFR(t)} \propto \rm{SFR}_{0} \times  \left \{
      \begin{array}{rcl}
          & 1 & t < 1\,\rm Myr,  \\ 
         & \exp \left[ \frac{-(t -1\,\rm Myr)^{2}}{t^{2}_{\rm d}} \right] & t \geq 1\,\rm Myr. 
      \end{array}
   \right .
    \label{eq:SFR}
\end{align}
This prescription yields a distribution of formation times, capturing the intrinsic age spread characteristic of stellar populations. This resulting distribution (shown in Fig.~\ref{Fig:SFR} in Appendix \ref{app:initial_dist}, functionally identical to \citealt{coleman2022dispersal}) considers a decay timescale of $t_{\rm d} =1$ Myr. Including this star formation history is crucial for distinguishing between intrinsic disc evolution and effects arising from the age spread of the population, particularly when reproducing the observed scatter in accretion rates and the disc fractions at older cluster ages \citep[e.g., in Upper Scorpius,][]{2021Squicciarini}. We discuss the specific impact of different SFR timescales in Sec.~\ref{sec:env-cond}.

\subsection{Initial disc mass}

The mass of the disc is dominated by the mass of the gas, which is mostly composed of molecular hydrogen and helium. While it is not possible to determine the total disc masses directly from molecular hydrogen emission, the thermal emission of the dust particles can be measured relatively easily and the measured flux values can be converted into dust masses \citep{tychoniec2018vla,manara2023ASPC}. We follow \cite{Emsenhuber_2021} which use the statistical distribution of gas disc masses inferred for protostars located in Perseus by \cite{tychoniec2018vla}. These authors convert the dust mass to gas mass using a typical gas-to-dust mass ratio of 100:1. This distribution represents a population of classes 0/I. The disc masses follow a log-normal probability distribution (shown in Fig. A.2 in Appendix A) where the mean value is $\log(M_{\rm d}/M_\star) = -1.49$ and the dispersion is $\sigma = 0.35$. A positive correlation between disc mass and stellar mass is well-established observationally for more evolved Class II discs \citep[see][and references therein]{Andrews2020ARAA}. 
In this work, we consider a sample limited to a $3\sigma$ range. This corresponds to a mass ratio $M_{\rm d}/M_\star$ from $\sim 0.003$ to $\sim 0.35$. To guarantee that all discs are gravitationally stable, we verified that all generated initial discs satisfy the Toomre criterion \citep[$Q > 1$,][]{Toomre1964ApJ...139.1217T} throughout their radial extent.

\subsection{Initial surface gas density}

The initial surface density is chosen according to the work of \cite{Andrews_2010}, based on the similarity solutions given by \cite{BellPringle1974} and \cite{Hartmann_1998}. Thus, the initial structure of the disc is represented by the following radial profile,
\begin{equation}
    \Sigma_{\rm g} = \Sigma_{\rm{gas}}^{0} \left(\frac{R}{{R}_{\rm c}}\right)^{-\gamma} e^{-(R/{R}_{\rm c})^{2-\gamma}}
    \label{equ:densi_inicial},
\end{equation}
where $\Sigma_{\rm{gas}}^{0}$ is a normalization constant related to the initial disc mass by 
\begin{equation}
        \Sigma_{\rm{gas}}^{0} = (2-\gamma) \frac{{M}_{\rm d, ~t=0}}{2\pi {{R}_{\rm c}}^2},
    \label{eq:Sigma_0_ a_T_0}
\end{equation}
being ${R}_{\rm c}$ the characteristic radius of the disc, $M_{\rm d}$ its initial mass and $\gamma$ the radial gradient which determines the initial mass distribution of the disc. For the surface density exponent we take the mean value obtained by \citet{Andrews_2010}, $\gamma = 0.9$, in all simulations.

For the characteristic radius $R_{\rm c}$ we follow \citet{Emsenhuber_2021}, who adopt the results from \cite{Andrews_2010}, and take $R_{\rm c}$ to be solely dependent on the mass of the disc,
\begin{align}
   \frac{R_{\rm c}}{10~\rm{au}} =   \left(\frac{M_{\rm{d}}}{2 \times 10^{-3} \rm{M}_{\odot}}\right)^{0.625}.
    \label{eq:rc_md}
\end{align}

\subsection{Disc evolution}
\label{sec:model}

Once the initial properties of the disc are defined, we need to model how different discs, around different stars, evolve with time. We compute the evolution of a circumstellar disc around a central star using the disc model implemented on {\scriptsize PLANETALP} \cite[e.g][]{pronco2017, OctavioMarcelo2017Codigo,  OGuilera2019MNRASb}. This model corresponds to an axi-symmetric 1D+1D, non-isothermal $\alpha$-disc model. The assumption of an $\alpha$-disc implies that the effective kinematic viscosity of the disc is given by  $\nu = \alpha c_{s}H_{g}$  \citep{shakura1973black}, where $H_{g}$ is the scale height of the disc, $H_{g} = c_{s}/\Omega$, where  $c_{s}$ is the local isothermal sound speed, and $\Omega$ is the keplerian-frequency. 

Within this framework the disc evolves by viscous accretion and photoevaporation, the latter both  due to the irradiation from the central star and external sources (e.g. young massive stars in the population). The evolution of the surface density of the gaseous disc $\Sigma_{\rm{g}}$ follows a diffusion equation given by \citep{pringle1981accretion}:
\begin{equation}\label{dens_sup_completa}
     \frac{\partial \Sigma_{\rm g} }{\partial t} =  \frac{3}{R} \frac{\partial}{\partial R} \left[  R^{1/2}\: \frac{\partial } {\partial R} \left(\nu \, R^{1/2} \, \Sigma_{\rm g}\right) \right] + \dot{\Sigma}_{\rm w}, 
\end{equation}
where $\dot{\Sigma}_{\rm w}$ represents the sink term due to photoevaporation \citep[e.g][]{OctavioMarcelo2017Codigo}. The time evolution of the surface density is solved using an implicit Crank-Nicholson scheme. We use 2000 radial bins logarithmically equally spaced that extend from the inner radius, $R_{\rm int}= 0.1 \, (M_{\star}/M_{\odot})^{1/3}$~au, to 1000 au \citep{Venturini+2024}.

For each value of the viscosity parameter $\alpha$, the relation $\nu(\Sigma_{\rm g},R)$ for a star of a given mass ($M_\star$) and effective temperature ($T_{\rm eff}$) is obtained from the resolution of the vertical structure of the disc taking into account the effects of the stellar irradiation \citep[see][for a detailed description of the method]{OctavioMarcelo2017Codigo}. Due to the computational cost of solving the non-isothermal disc to perform a population synthesis with our model we implemented a grid of our vertical model in stellar mass and $\alpha$ parameter. We include stellar masses in the range 0.04 to 1.4 $\rm M_{\odot}$. We sample the continuum by taking intervals of $0.01~\rm M_{\odot}$ in the range 0.04 -- 0.1~$\rm M_{\odot}$, and $0.1~\rm M_{\odot}$ in the range 0.1 -- 1.4~$\rm M_{\odot}$. The parameter $\alpha$ takes discrete values in the range $10^{-5.5}$-$10^{-1}$. We considered ten linearly spaced values in $\log \alpha$. To obtain the relation $\nu(\Sigma_{\rm g},R)$ for arbitrary values of $\alpha$ and $M_{\star}$, we use a bilinear interpolation in $\log \alpha$ and $\log M_{\star}$.

\subsubsection{Internal and external photoevaporation}

Photoevaporation plays a crucial role in the dispersal of protoplanetary discs. In our model, we account for both internal photoevaporation driven by the central star and external photoevaporation driven by the surrounding environment. The total mass-loss sink term incorporated into the diffusion equation is given by $\dot{\Sigma}_{\rm w} = \dot{\Sigma}_{\rm XR} + \dot{\Sigma}_{\rm FUV} + \dot{\Sigma}_{\rm ext~fuv}$.

Internal photoevaporation is driven by X-ray (XR) and far ultraviolet (FUV) radiation from the central star. The mass loss due to XR photoevaporation is computed following the analytical prescriptions by \cite{owen2012theoryfoto}, which strongly depends on the stellar XR luminosity and drives a thermal wind primarily from the inner disc. The rate of FUV photoevaporation is incorporated following \citet{Kunitomo2021}. For a detailed description of the numerical implementation of these internal mass-loss rates in the PLANETALP code, we refer the reader to \cite{OctavioMarcelo2017Codigo} and \cite{Venturini+2024}.

Furthermore, in star-forming regions, massive stars dominate the production of UV radiation, leading to external photoevaporation that efficiently truncates the disc from the outside-in. Following \cite{coleman2022dispersal}, disc mass is removed from the outer regions beyond the gravitational radius. To explore the impact of different environments, we adopt parameterized mass-loss rates for a reference disc of 100~au, $\dot{M}_{\rm d}^{fuv,0}$, ranging from $10^{-9}$ to $10^{-7} \, M_{\odot} \, \text{yr}^{-1}$. These values are representative of discs exposed to a moderate UV field of $G_0 = 10$ \citep[see Table 2 in][]{haworth2018fried, anania2025a}.

The explicit recipes used to compute both the internal and external photoevaporation profiles in our simulations are detailed in Appendix \ref{app:photoevaporation}.

\subsubsection{Disc viscosity}

Within the $\alpha$-disc framework \citep{shakura1973black}, the macroscopic kinematic viscosity $\nu$ is determined by the dimensionless parameter $\alpha$. This parameter quantifies the degree of turbulence in the disc and is empirically calibrated to reproduce the observed accretion rates and disc lifetimes \citep[e.g.][]{2020A&A.A.R}. However, there is still no consensus on the most appropriate values of $\alpha$ when modelling protoplanetary discs. Values in the range $10^{-4}$–$10^{-1}$ are typically required to reproduce the observed accretion rates \citep[see][for a discussion]{2020A&A.A.R}.  

Observations also show that the stellar accretion rate is correlated with the stellar mass, approximately following $\dot{M} \propto M_{\star}^{2}$ \citep[see][and references therein]{Gorti_2009, Dullemond2006, Alexander2006}. It remains an open question whether this correlation is a consequence of disc evolution or merely reflects how the initial conditions scale with stellar mass \citep[e.g.][]{2017Ecol, manara2023ASPC}. Following \cite{Gorti_2009}, we assume that the correlation arises from the dependence of the turbulent properties of the disc on stellar mass, such that $\alpha = \alpha(M_{\star})$.  

To account for the observed dispersion in the accretion rates at a given stellar mass, we follow \cite{manara2023ASPC} and introduce a dispersion in $\alpha$ by means of an auxiliary parameter $\alpha_0$, such that  
\[
\alpha = \alpha_{0} \times M_{\star}^{a},
\]  
where $a$ is a free parameter to be calibrated. We explore the effect of adopting two different distributions for $\alpha_0$: a uniform distribution and a linear one. In the latter case, for a given stellar mass, more viscous discs are more likely to occur around higher-mass stars.  

\begin{figure}\centering
\includegraphics[width=\hsize]{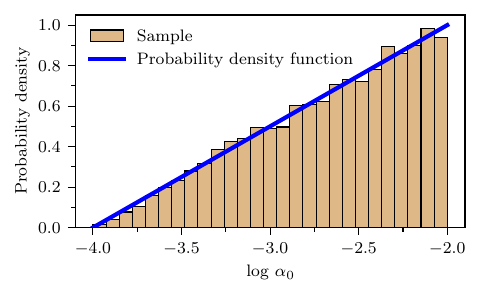}
    \caption{Linear probability distribution for $\log \alpha_{0}$ between -4 and -2. This distribution favours the presence of high viscosity discs in our population synthesis.}
    \label{fig:refdistrlineal}
\end{figure}

Regarding the range of $\alpha_0$, we consider two cases: $\log \alpha_0 \in [-4.0,-2.0]$ and $\log \alpha_0 \in [-3.5,-1.5]$. Fig.~\ref{fig:refdistrlineal} illustrates an example of the linear probability density function for $\log \alpha_{0}$. In this scenario, high-viscosity discs are favoured—rather than being uniformly distributed—for a given stellar mass.

\section{Observational constraints on disc evolution models}
\label{sec:constrains}

 Observations of protoplanetary discs provide important constraints on their properties and evolution. The fraction of stars with discs as a function of the young stellar cluster ages, the masses of the discs, their dissipation timescales, and the accretion rates onto the central stars are some of the most important observational constraints currently available. 

For this work, we consider the statistical sample obtained by \cite{manara2023ASPC} of the stellar mass accretion rates as a function of the disc masses and as a function of the stellar masses for almost 1000 young stellar objects within 300 pc. The sample contains young stellar clusters at different evolutionary stages and environments. The stellar mass accretion rates as a function of the disc masses are inferred from the dust disc masses from \cite{2019A&A.Manara} using the standard dust-to-gas ratio of 1\% similar to that of the interstellar medium. In Sec.~\ref{sec:uncertaintyMd} we analyse the impact of uncertainty in the dust-based gas mass estimates.

In addition, to constrain the lifetime of the protoplanetary discs, we considered the result from \cite{Pfalzner_2022}. The observations of clusters closer than 200 pc, wherein all the stellar mass distribution can be well estimated, allow one to obtain median disc lifetimes between 5 -- 10 Myr. Clusters at larger distances seem to underestimate the contribution of low-mass stars, deriving short disc lifetimes of about 1 -- 3 Myr. The main reason for this discrepancy is that distant clusters are affected by a limiting magnitude, thus in the samples of distant clusters low-mass stars are under-represented. In order to compare the observed fraction of stars with discs depending on the cluster distances \citep{Pfalzner_2022}, we implement in our population synthesis a cut-off in our mass distribution, mimicking a limiting magnitude due to the cluster distances.  

\section{Results}
\label{sec:result}

We present the results of our simulations by comparing synthetic disc population models with available observational data. Our analysis focuses on disc lifetimes as a function of host star properties and local formation conditions. In addition, for each synthesis, we extract key observables: disc lifetimes inferred from the disc-star fraction over the 1–20 Myr age range, and mass accretion rates onto the central star as functions of both disc mass and stellar mass at 1.5 Myr --a representative age for young clusters such as Taurus, Lupus, and Chamaeleon I \citep{manara2023ASPC}. In order to compare the simulations against the observations, we compute the mass accretion onto the star as
\begin{align}
    \dot{M} = 2\pi \mathbf{R_{\rm in}} \Sigma_{\rm g}(\mathbf{R_{\rm in}}) \, v_{\rm gas}(\mathbf{R_{\rm in}}),
    \label{eq:synthetic_accretion}
\end{align} 
$v_{\rm gas}(R)$ being the gas radial velocity \citep[see][]{Garate2020A&A}. We note that for comparison, $\dot{M}$ is computed in the innermost part of the disc, evaluated at the inner radius $\mathbf{R_{\rm in}}$. 

To assess the impact of the limiting detection magnitude on the inferred mean disc dissipation times --derived from the fraction of stars with discs as a function of cluster age--, we performed a series of simulations using different lower limits for the stellar mass distribution. To analyse the effect of adopting a simplified SFR, we considered two scenarios: one in which all stars form simultaneously within the cluster, and the other in which star formation follows a temporal distribution characterized by a timescale of $\rm{t}_{\rm{d}} = 1$~Myr.

\subsection{Reference synthesis}

We first generate a population of 10000 star–disc systems, with model parameters drawn from distributions inferred from observations, as described in the previous section. In this reference synthesis, we assume the $\alpha$-viscosity parameter follows a uniform distribution in logarithmic space between $\log \alpha= -4$ and $\log \alpha = -2$, consistent with typical values used to reproduce observed disc properties \citep[e.g.][]{2020A&A.A.R, 2023Rosotti}.

\begin{figure*}[!!ht]
\centering
\includegraphics[width=\textwidth]{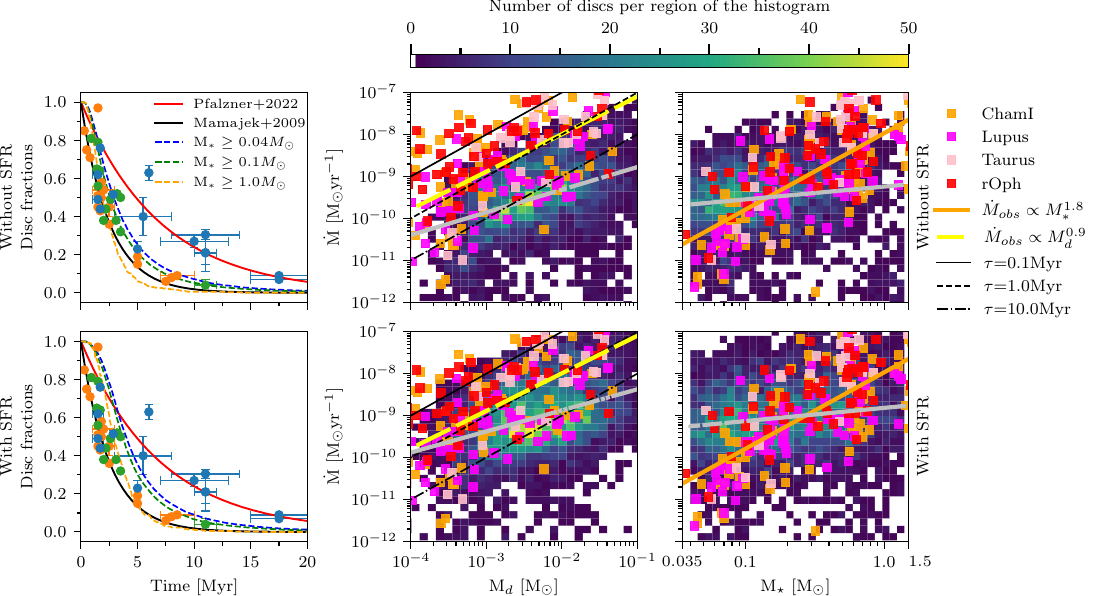}
\caption{Top panels. Disc population synthesis results without considering a prolonged SFR. Bottom panels. Results adopting an SFR with a characteristic timescale $t_{\rm d}= 1$~Myr. Left column. Fraction of stars with discs as a function of the age of the cluster. Blue, green and orange points correspond to observations of young stellar clusters within 200~pc, between 200--500~pc and between 500-1000~pc, respectively. The red line represents the fit to cluster within 200~pc \citep{Pfalzner_2022}. The black line corresponds to the results obtained by \cite{MamajekE}. The dashed lines represent the results of our population synthesis for different values of $M_{\star,\rm m}$ considered. The orange, green and blue dashed lines correspond to $M_{\star,\rm m}=1,\,0.1$ and 0.04$ \rm M_{\odot}$, respectively. Center column. Histogram (or colour density map) for the stellar mass accretion rate as a function of the gas disc mass at 1.5 Myr from the population synthesis. Orange, magenta, pink, and red points represents the observations compiled in \citet{manara2023ASPC}. Right column. Stellar mass accretion rate against stellar mass, also at 1.5 Myr of the synthesis evolution. The orange line represents the fit to the observations \citep{manara2023ASPC}, while the gray line is the fit to our results.} 
\label{Fig:reference}
\end{figure*}

The main results of the reference synthesis are shown in Fig.~\ref{Fig:reference}. The top row displays the results assuming all star–disc systems begin their evolution simultaneously at time zero (i.e., without considering a prolonged SFR), while the bottom row presents the same results incorporating a time-dependent SFR.

The left column of Fig.\ref{Fig:reference} shows the fraction of stars with discs as a function of cluster age. The data points represent observed disc fractions with their associated uncertainties. Blue points correspond to clusters within 200 pc, green to those between 200–500 pc, and orange to clusters between 500–1000 pc \citep{Pfalzner_2022}. The red line shows the disc fraction fit for a low-mass star-corrected sample from \citet{Pfalzner_2022}, while the black line shows the classical fit from \citet{MamajekE}. The latter may overestimate the contribution of massive stars, as it does not account for cluster distance effects \citep[see][]{Pfalzner_2022, 2024ApJPfalzner}.

Our synthetic results are shown as dashed lines. To mimic the varying stellar mass completeness across clusters at different distances, we apply different lower limits on the stellar mass when computing disc fractions. The blue dashed line represents the full synthetic population, spanning stellar masses from $0.04~\rm{M}_{\odot}$ to $1.4~\rm{M}_{\odot}$. Notably, we observe a steep decline in disc fractions at early ages, yielding a mean synthetic disc lifetime of $\tau =3.99 \pm 0.01$~Myr without SFR 
(see Table \ref{tab:lifetimes} in Appendix \ref{app:tables}), which underestimates the disc fractions in clusters where low-mass stars are well sampled (cf. red line).

When applying higher mass cut-off ($M_{\star,\rm m}$) to reflect observational biases at greater distances --specifically, 0.1 and 1.0~$\rm{M}_{\odot}$ for the green and orange dashed lines, respectively--, we observe an even lower disc fraction, consistent with the faster disc dispersal around massive stars \citep[e.g.][]{ribas2015protoplanetary}. A mass threshold of $1~\rm{M}_{\odot}$ is required to reproduce the disc fractions reported by \citet{MamajekE} and those observed in distant clusters (500 to 1000 pc) \citep{Pfalzner_2022}.

When the SFR is included in the calculation of the fraction of stars with discs (bottom-left panel of Fig.~\ref{Fig:reference}), we observe an increase in the number of disc-bearing stars at young cluster ages and the mean synthetic disc lifetime reaches  $\tau = 5.08 \pm 0.01$~Myr. The inclusion of the SFR yields disc lifetimes consistent with the results reported by \citet{coleman2022dispersal}. This increase is attributed to ongoing star formation during the first few million years, which populates the clusters with newly formed stars still hosting protoplanetary discs.

The central panels of Fig.~\ref{Fig:reference} show a colour density map (i.e. the number of systems per pixel) of the stellar mass accretion rate as a function of the gas disc mass when the age of the cluster is 1.5~Myr. The orange, magenta, pink and red points correspond to the observed disc masses in the star-forming regions of Chamaeleon I, Lupus, Taurus and Ophiuchus, respectively, and considering a classical dust-to-gas ratio of 1\% \citet{manara2023ASPC}. The dashed-dot lines represent different dispersal timescales, $\tau= M_{\rm d}/ \dot{M}$, of 0.1, 1 and 10 Myr. Without considering an SFR (top-central panel), our synthesis is unable to reproduce the observed high accretion rates. In order to better reproduce the observations for low-mass discs and high accretion rates, it is necessary to introduce a prolonged SFR. The inclusion of an SFR affects the accretion rate--disc mass relation by shifting systems above the dissipation timescale curves $\tau = M_{\rm d}/\dot{M}$. This shift is primarily driven by the continuous formation of young star--disc systems during the first few million years, which introduces low-mass discs with relatively high accretion rates. As a result, the inclusion of the SFR increases the scatter in both the $\dot{M}$--$M_{\rm d}$ and $\dot{M}$--$M_{\star}$ correlations. This additional dispersion tends to improve the agreement between the synthetic accretion rates and those observed in young stellar populations. However, we are still unable to match the observed accretion rates for low-mass discs.  

The right column of Fig.~\ref{Fig:reference} displays the stellar mass accretion rate as a function of the stellar mass. Observational data from young stellar clusters are shown again as orange, magenta, pink, and red points. The orange line represents the observational fit, $\dot{M}_{\rm obs} \propto M_{\star}^{1.8}$, as reported by \citet{manara2023ASPC}. The corresponding trend from our synthetic population is shown as a gray curve. Due to the overproduction of low-mass stars with high accretion rates in our models, the resulting correlation is significantly flatter than observed. Our linear regression yields a slope of $a_1 = 0.315 \pm 0.0003$ for the model with SFR (Table \ref{tab:regresiones_pendientes_mstar} in Appendix \ref{app:tables}), failing to reproduce the steeper empirical correlation of $a_1 = 1.823 \pm 0.007$ \citep{manara2023ASPC}.

\subsection{Disc viscosity as a function of the stellar mass}
\label{sec:res-corr}

The results of our reference synthesis presented in Section 4.1 indicate the need for a correlation between disc viscosity and stellar mass. Observations support this, showing $\dot{M}{\rm obs} \propto M_{\star}^{1.8}$ \citep{manara2023ASPC}. We therefore calculate new population syntheses considering a correlation for the $\alpha$-viscosity parameter of the form $\alpha = \alpha_{0} \, M_{\star}^{a}$. We also consider two different distributions for $\alpha_{0}$: first we adopt that $\log \alpha_0$ follows a uniform distribution; then we assume a linear distribution (see previous section). We consider three values for the stellar mass exponent $a= 1.5, 1.8$, and 2.0.  

\subsubsection{$\log \alpha_{0}$ between -4 and -2}
\label{sec:res-corr1}

In Appendix \ref{sec:appendix_viscosity} (Fig.~\ref{fig:LR_app}), we show the results of our synthesis considering $\log \alpha_0 $ in the range -4 to -2 and the value of $a = 1.8$ (the results for the syntheses adopting $a = 1.5$ and 2 are also detailed in Appendix \ref{sec:appendix_viscosity}). The top panels show the results of considering the uniform probability distribution for $\log \alpha_0$, while the bottom panels show the result of considering the proportional probability distribution for $\log \alpha_0$. In both cases we consider an SFR. 

Including a viscosity dependence on stellar mass results in synthetic disc fractions that closely match the estimations of \citet{Pfalzner_2022}. The mean synthetic disc lifetime reaches $\tau = 8.16 \pm 0.02$~Myr for the uniform distribution and $\tau = 7.78 \pm 0.02$~Myr for the linear distribution (Table \ref{tab:lifetimes}). When we apply a mass cut-off at $1~\rm M_{\odot}$,  to mimic in some way the observational biases at larger cluster distances, our synthetic disc fractions also reproduce the disc lifetimes inferred by \cite{MamajekE}, and the observed disc fractions of young stellar clusters with distances larger than 500~au.

The accretion rate as a function of the disc mass is shown in the central column of Fig.~\ref{fig:LR_app}. Due to the correlation between $\alpha$ and stellar mass, we can see that low- and moderate-mass discs show a decrease in the mass accretion rates, not being able to reproduce the observations.  

The accretion rate as a function of the stellar mass is shown in the right column of Fig.~\ref{fig:LR_app}. By adopting a stellar mass dependence for the disc viscosity, both distributions for $\alpha_0$ (uniform and linear) reproduce the $\dot{M}- M_{\star}$ correlation inferred by observations, with the linear distribution of $log \alpha_0$ leading to a better fit. However, it is evident that the synthetic accretion rates are considerably lower than those estimated by the observations, with a high degree of dispersion in the values.

\subsubsection{$\log \alpha_{0}$ in the range -3.5 to -1.5}
\label{sec:-3.5-1.5}

\begin{figure*}[!!ht]
    \centering
    \includegraphics[width=\textwidth]{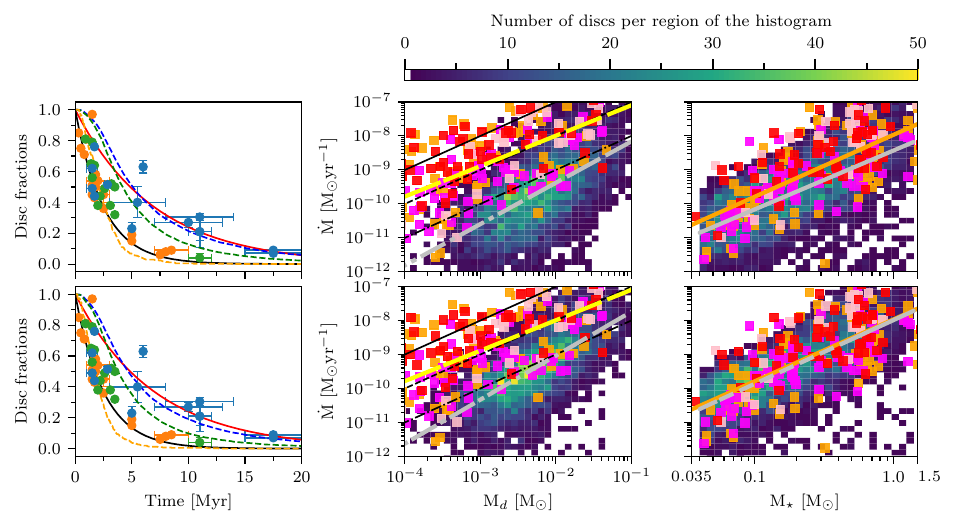}
    \caption{Same as Fig. \ref{Fig:reference} (although exploring a different $\log \alpha_0$ range), but considering the range of $\log \alpha_0$ between -3.5 and -1.5.}
    \label{fig:hi}
\end{figure*}

In the previous section we showed that if the disc viscosity is proportional to the stellar mass, the trend in the observed correlation $\dot{M}- M_{\star}$ can be recovered. However, the synthetic accretion rates are smaller than the observed ones. With the aim of obtaining larger synthetic accretion rates from our population, we examine a new synthesis adopting a higher range for $\log\alpha_{0}$. In particular, we adopt a range between -3.5 and -1.5 for $\log\alpha_{0}$. As in the previous section, we consider two population synthesis, one following a uniform distribution, and the other assuming a linear distribution between -3.5 and -1.5. The results considering $a=1.8$ are shown in Fig.~\ref{fig:hi}. The results for the syntheses adopting $a = 1.5$ and 2 are detailed in App.~\ref{sec:appendix_viscosity}. Despite the increment in the range of the values for the $\alpha_0$-viscosity parameter, we note that since the distribution of $\alpha$ is weighted by the stellar mass the effective range of $\alpha$ is between $\sim 5 \times 10^{-5}$ and $\sim 6 \times 10^{-2}$.

The left panels of Fig.~\ref{fig:hi} show that our results satisfactorily reproduce the fraction of stars with discs for clusters with distances less than 200~pc. The synthetic populations yield mean disc lifetimes of $\tau = 7.15 \pm 0.01$~Myr for the uniform distribution and $\tau = 6.57 \pm 0.01$~Myr for the linear one (Table \ref{tab:lifetimes}). Also in both cases, considering a limiting stellar mass of $1\rm M_{\odot}$ for the population allows us to reproduce the results obtained by \cite{MamajekE}. 

In the right panels of Fig.~\ref{fig:hi}, we plot again the stellar mass accretion rate as a function of the star mass. We find that the increment in the values of $\log \alpha_0$ allows both distributions to reproduce very well the observational trend of $\dot{M} - M_{\star}$. In particular, the linear distribution for $\log \alpha_0$ yields an excellent quantitative agreement with the observations in the $\dot{M}-M_{\star}$ plane. The synthetic population produces a linear regression slope of $a_1 = 1.875 \pm 0.0003$, successfully recovering the observed empirical slope of $a_1 = 1.823 \pm 0.007$ (Table \ref{tab:regresiones_pendientes_mstar}).

However, as seen in the central panels of Fig.~\ref{fig:hi}, we still fail to reproduce the stellar mass accretion as a function of the gas disc mass. None of the synthetic populations are able to efficiently generate low- and intermediate-mass discs with high accretion rates. We note however, that one possibility to better reproduce these observational estimations is to consider a phenomenon ---as external photoevaporation--- that could reduce gas disc masses maintaining high accretion rates. Another possibility is to  consider that, at 1.5~Myr, a significant fraction of solids is locked in larger pebbles, planetesimals, or even planets, not contributing to the infrared observations, and hence the inferred gas masses can be underestimated. We analyse these two possibilities in the following sections.

\subsection{The impact of environmental conditions}
\label{sec:env-cond}

The correlations between stellar mass accretion rates and disc masses arise naturally within the framework of viscous disc evolution. In this scenario, discs start as massive structures with high accretion rates that decrease as they evolve. However, models relying solely on viscous evolution and internal photoevaporation often struggle to reproduce the full spread of mass accretion rates, particularly for low-mass discs. Specifically, internal photoevaporation opens an inner hole when the photoevaporation mass-loss rate is comparable to the viscous accretion rate, effectively cutting off accretion to the central star \citep{clarke2001, owen2011, Somigliana2020}. This mechanism tends to deplete low-mass discs with accretion too quickly or creates a population of non-accreting discs, which contradicts observations of young star clusters that show low-mass discs with high accretion rates \citep{Somigliana2020, emsenhuber2023towards}.

 In contrast, when external photoevaporation is included, the evolutionary pathway changes significantly. External photoevaporation removes material efficiently from the outer disc regions beyond the gravitational radius \citep{haworth2018fried, coleman2022dispersal}. Consequently, the internal part of the disc remains rich in material and continues to evolve mainly by viscosity, maintaining high accretion rates onto the star even as the total disc mass decreases. This mechanism allows the model to reproduce the population of low-mass discs with high accretion rates that purely internal photoevaporation models fail to capture.

\begin{figure*}[!!ht]
    \centering
    \includegraphics[width=\textwidth]{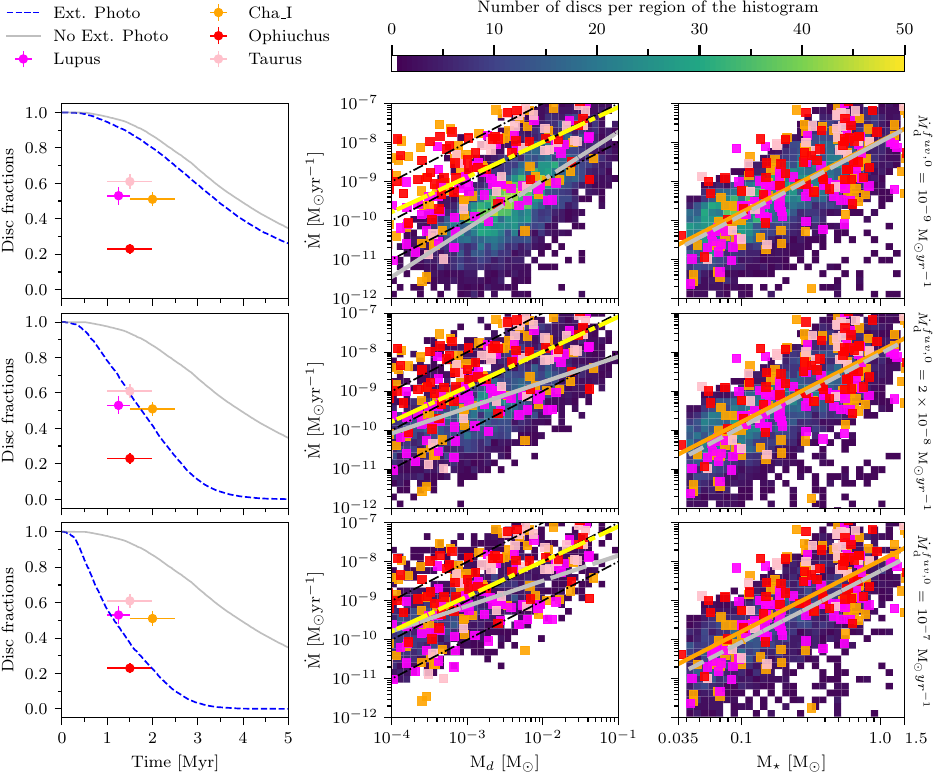}
    \caption{Population synthesis results for three different values of $\dot{M}_{\rm d}^{\rm fuv,0}$ and an SFR with $t_{\rm d}= 1$~Myr. From top to bottom, the rows correspond to $\dot{M}_d^{\rm fuv,0}  = 10^{-9}$, $2 \times 10^{-8}$, and $10^{-7} \, \rm M_{\odot} yr^{-1}$. Left column: fraction of stars with discs as a function of cluster age. The solid grey lines represent the reference model without external photoevaporation, while the dashed blue lines include this mechanism. Middle column: stellar mass accretion rate versus gas disc mass at 1.5 Myr. Right column: stellar mass accretion rate versus stellar mass at 1.5 Myr. Observational data from Lupus, Chamaeleon I, Taurus, and Ophiuchus are taken from \cite{manara2023ASPC}.}
\label{fig:WExtrnalPhotoevaporation}
\end{figure*}

Thus, we compute new disc population syntheses including external photoevaporation with the aim to also reproduce the observed trend for the accretion rate as a function of the disc mass. We consider that $\log \alpha_0$ follows a linear distribution in the range between -3.5 and -1.5, adopting the viscosity scaling $a= 1.8$, and an SFR with a characteristic decay time of $t_d = 1$ Myr. The results are presented in Fig.~\ref{fig:WExtrnalPhotoevaporation}. The top panels show the result for $\dot{M}_{\rm d}^{\rm fuv,0}=10^{-9} M_{\odot} \, \text{yr}^{-1}$, middle panels show the cases for $\dot{M}_{\rm d}^{\rm fuv,0}=2 \times 10^{-8} M_{\odot} \, \text{yr}^{-1}$ and the bottom panels for $\dot{M}_{\rm d}^{\rm fuv,0}=10^{-7} M_{\odot} \, \text{yr}^{-1}$. As before, the left panels display the evolution of the disc fraction as a function of cluster age. The blue dashed lines represent the full synthetic population ($0.04 \le M_{\star}/{\rm M}_{\odot} \le 1.4$) evolving under the combined influence of viscous evolution, internal and external photoevaporation, and considering the star formation rate previously mentioned. For comparison, the grey dashed lines depict the synthesis not considering external photoevaporation. Observational estimates for the young stellar clusters of Lupus, Chamaeleon I, Ophiuchus, and Taurus, are shown for comparison \citep{ribas2015protoplanetary, Pfalzner_2022}. The central and right panels of Fig.~\ref{fig:WExtrnalPhotoevaporation} show the $\dot{M}$--$M_{\rm d}$ and the $\dot{M}$--$M_{\star}$ planes, respectively.

We find that the best match to the observations is achieved by the model with $\dot{M}_{\rm d}^{\rm fuv,0} = 2 \times 10^{-8} \, M_{\odot} \, \text{yr}^{-1}$. This model yields a mean disc lifetime of $\tau = 1.92 \pm 0.01$~Myr (Table \ref{tab:lifetimes}), reproducing the observed decay in disc fractions without over-predicting long-lived discs. Furthermore, this synthetic population reproduces the $\dot{M}-M_{\star}$ correlation slope robustly ($a_1 = 1.879 \pm 0.0004$, Table \ref{tab:regresiones_pendientes_mstar}), confirming that external photoevaporation does not disrupt the stellar mass dependency.

Our synthesis reproduces again very well the observed correlation $\dot{M}_{\rm obs} \propto M_{\star}^{1.8}$ (including the magnitudes and dispersion of the observed accretion rates).  

Thus, to reproduce the observed accretion rates in young clusters, it is necessary to consider the effect of the environment produced by the external photoevaporation  and a population formed via a star formation rate adopting a decay time of 1~Myr. However, in the case of discs with high accretion rates at extended ages, as is the case of the Upper Scorpius cluster, it is necessary to consider an extended star formation rate (which is supported by the observations, see discussed in the Upper Scorpius paragraph.).

\subsubsection*{The interesting case of Upper Scorpius}

In this section, we look for the initial conditions of our model that allow us to reproduce the observable characteristics of the young stellar cluster Upper Scorpius. This cluster appears to be older than those previously studied, with an estimated age between $\sim 5$~Myr and $\sim 11$~Myr \citep{fang2017, 2021Squicciarini, manara2023ASPC}. Upper Scorpius presents a non-negligible number of components with high accretion rates despite its estimated age, with a high $\dot{M} - M_{\rm d}$ dispersion.

\begin{figure*}
    \centering
    \includegraphics[width=\textwidth]{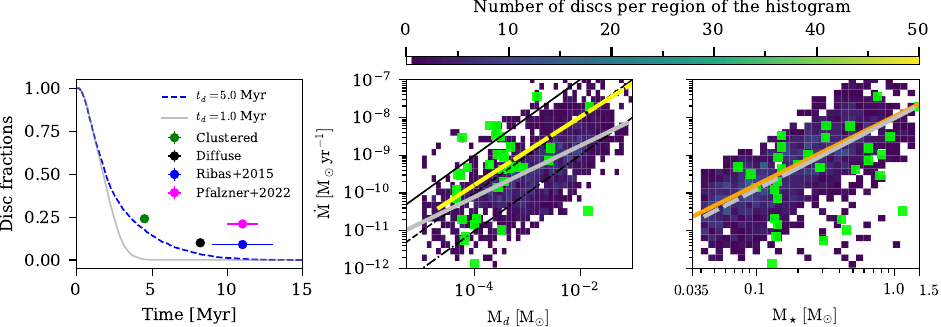}
    \caption{Same as  middle panels of  Fig.~\ref{fig:WExtrnalPhotoevaporation} but adopting an extended star formation rate with $t_{\rm d}= 5$~Myr. For comparison, the grey line in the left panel corresponds to the synthesis results adopting $t_{\rm d}= 1$~Myr. The green dots correspond to the observational data from \citet{manara2023ASPC} for the Upper Scorpius cluster. The colour dots represent our synthetic accretion rates at 11~Myr of the synthesis evolution.}
    \label{fig:UpperS}
\end{figure*}

In order to reproduce observations, we need to consider a large characteristic time of the star formation rate of 5~Myr, and as in Sec. \ref{sec:env-cond} we consider that the simulations are affected by external photoevaporation (in order to reproduce low-mass discs with high accretion rates). Fig.~\ref{fig:UpperS} represents the results of our synthesis adopting a value of $\dot{M}_d^{\rm fuv,0}  = 2 \times 10^{-8}$. The central and right panels represent the stellar mass accretion rates as a function of the disc masses and as a function of the star masses, respectively, at 11~Myr of the synthesis evolution. 

Our results show that substantial accretion rates can still be present in an evolved cluster such as Upper Scorpius if a prolonged star formation history is taken into account. This setup yields a mean disc lifetime of $\tau = 2.691 \pm 0.004$~Myr (Table \ref{tab:lifetimes}), consistent with the clustered and diffuse components given by \citet{2021Squicciarini}. Conversely, if a star formation history is not considered, our simulations show that by 11~Myr, all discs have dissipated and no accreting stars remain in the population. In addition, if FUV external photoevaporation is also not considered, we can not account for low-mass discs with high accretion rates.

The left panel of Fig.\ref{fig:UpperS} shows the fraction of stars with discs as a function of time. We can see that our synthesis also matches well the observed disc fraction estimations \citep{ribas2015protoplanetary, Pfalzner_2022}. As we mentioned before, the age of Upper Scorpius remains uncertain. While classical estimates place it at $\sim$5 Myr, recent studies using 3D kinematics and precise stellar modeling suggest older ages of $\sim$10--12 Myr \citep{Ratzenbock2023}. \citet{2021Squicciarini} suggest that the discrepancy in the age estimation could be partially explained by an intrinsic age spread within the region. Such age spread could account for the relatively high number of accreting stars observed, in agreement with our results. In addition, our synthetic fraction of stars with disc is closer to the estimations for the clustered and diffuse components given by \citet{2021Squicciarini}.

\subsection{Uncertainty in the dust-based gas mass estimates}
\label{sec:uncertaintyMd}

 In the previous sections, we estimated gas disc masses from sub-mm continuum observations assuming that the observed mm emission traces the total solid reservoir, together with a standard interstellar gas-to-dust ratio of 100. However, this assumption is subject to significant uncertainties. Dust evolution processes such as grain growth and radial drift can efficiently deplete the population of mm-sized grains, which dominate the continuum emission, without necessarily reducing the total solid mass of the disc \citep{Sellek2019, Rosotti2019MN}. As a consequence, the dust mass inferred from mm observations may represent only a fraction of the total solid reservoir.

 As discussed in Sec.~4.2, even when adopting a stellar-mass-dependent viscosity that reproduces the $\dot{M}-M_\star$ relation, our models still fail to account for the population of discs with low inferred masses and high accretion rates. In Sec.~4.3, we showed that external photoevaporation provides a viable mechanism to explain this discrepancy. However, an alternative ---or complementary--- explanation is that gas disc masses inferred from dust continuum emission are systematically underestimated.

 This effect can be interpreted in terms of a visibility fraction, $f_{\rm vis}$, defined as the fraction of the total solid mass traced by mm observations. The standard assumption corresponds to $f_{\rm vis}=1$, whereas $f_{\rm vis}<1$ implies that a significant fraction of the solids is locked in larger bodies such as pebbles, planetesimals, or even planets, which are inefficient emitters at mm wavelengths.

 Recent observational and theoretical works support this scenario. By analysing CO isotopologue rotation curves in a sample of transition discs, \cite{longarini2025} derived dynamical gas masses that, when compared to continuum-based dust masses, imply effective gas-to-dust ratios of $\approx 400$, significantly higher than the canonical value of 100. In addition, \cite{savvidou2025} showed that pressure bumps induced by forming planets can efficiently trap solids, leading to a large fraction of the mass being hidden in non-emitting components.

\begin{figure}[!!ht]
    \centering
    \includegraphics[width=\hsize]{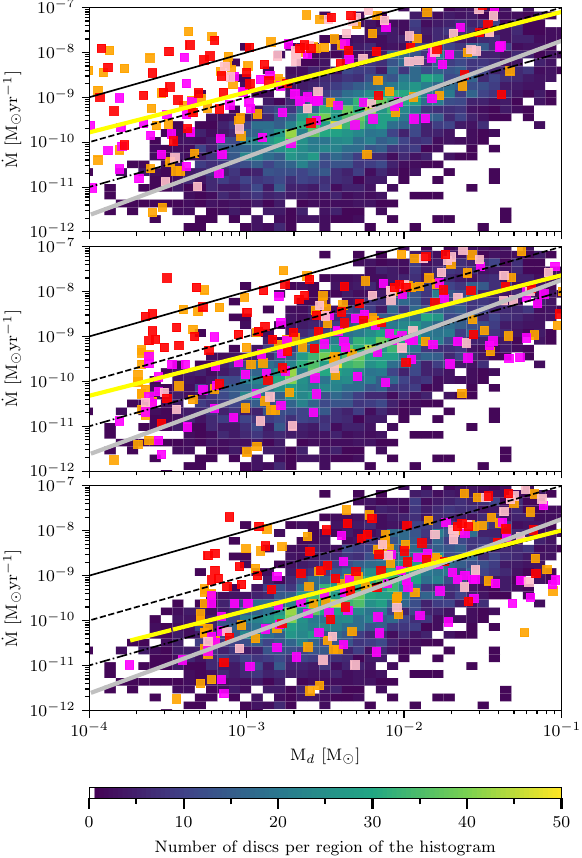} 
    \caption{ $\dot{M}-M_{\rm d}$ diagrams for the population synthesis adopting $a=1.8$ and a linear distribution of $\log \alpha_0$ between $-3.5$ and $-1.5$. Disc masses are inferred from dust continuum observations using different conversion factors between the observed dust mass and the total gas mass. From top to bottom, the panels show scaling factors of 100, 400, and 1000. These values can be interpreted as cases where $\sim 100\%$, $25\%$, and $10\%$ of the total solid mass is traced by mm emission, respectively. Increasing the conversion factor shifts the population horizontally towards higher inferred disc masses, resulting in longer accretion timescales $t_{\rm acc}=M_{\rm d}/\dot{M}$.}
    
    \label{fig:MdotMdVarGDFrac}
\end{figure}

 Motivated by these results, we explore the impact of increasing the conversion factor between the observed dust mass and the total gas mass in the $\dot{M}-M_{\rm d}$ plane. Figure~\ref{fig:MdotMdVarGDFrac} presents the results of our population synthesis (with $a=1.8$ and $\log \alpha_0 \in [-3.5, -1.5]$) adopting different scaling factors relative to the standard assumption used by \cite{manara2023ASPC}. The left panel corresponds to the canonical factor of 100, while the panels to the right consider larger factors of 400 and 1000. These values can be interpreted as cases where only 25\% and 10\% of the total solid mass is traced by mm emission, respectively; the extreme case of $f_{\rm vis}=0.1$ (ratio 1000) is included as a theoretical upper boundary to illustrate a scenario where the vast majority of solids are hidden in planetesimals or trapped in strong pressure bumps, as recently modelled by \cite{savvidou2025}.

Increasing this conversion factor shifts the synthetic population horizontally towards higher inferred disc masses, while leaving the accretion rates unchanged. As a consequence, the inferred accretion timescale, $t_{\rm acc} = M_{\rm d}/\dot{M}$, increases proportionally. Under the standard assumption, a significant fraction of discs appears to have short accretion timescales ($<1$ Myr), shorter than typical disc lifetimes. In contrast, adopting larger conversion factors shifts the population towards $t_{\rm acc} \sim 1$--$10$ Myr, bringing the inferred evolution into better agreement with viscous disc evolution expectations.

This result suggests that part of the observed scatter in the $\dot{M}-M_{\rm d}$ plane, particularly the population of discs with high accretion rates and low continuum-inferred masses, may be explained by a systematic underestimation of gas masses due to dust evolution and trapping. In this context, mm continuum emission should be regarded as a tracer of the observable dust reservoir rather than of the total disc mass, providing an alternative or complementary explanation to external photoevaporation.

\section{Discussion}
\label{sec:discussion}

Our population synthesis models explore how initial conditions, stellar-mass-dependent viscosity, and environmental factors like external photoevaporation affect disc lifetimes and accretion rates. Furthermore, we demonstrated that uncertainties in gas disc mass estimates due to dust evolution can act as a degenerate mechanism alongside photoevaporation to explain the spread in the $\dot{M}-M_{\rm d}$ plane. Below, we contextualize our findings within the broader landscape of disc evolution models and discuss their implications for planet formation.

\subsection{Comparison with previous studies}
\label{sec:disc_comparison_previous_works}

Our results reinforce and expand upon several previous findings regarding protoplanetary disc evolution. One key point is the observed large scatter in the correlation between mass accretion rate and stellar mass. This dispersion has been long noted in previous works \citep[e.g.][]{manara2023ASPC}, and our model successfully reproduces it by introducing a distribution in the $\alpha$-viscosity parameter $\alpha_0$, as well as a dispersion in the initial disc mass distribution. The synthetic mass accretion rate of our model, given by Eq.~(\ref{eq:synthetic_accretion}), links the mass accretion rate to both the surface density and the gas radial velocity, which are in turn controlled by $\alpha$-viscosity parameter. Without a dependence of $\alpha$ on stellar mass, our simulations predict a nearly flat $\dot{M} - M_{\star}$ relation (as in Fig.~\ref{Fig:reference}), which underestimates the observed correlation. However, by introducing a scaling of the form $\alpha \propto M_{\star}^a$, with $a= 1.8$, we are able to recover the observed trend \citep{Gorti_2009, manara2023ASPC}. Additionally, our population synthesis results confirm the importance of observational biases due to limiting magnitudes, as highlighted by \citet{Pfalzner_2022}. These biases lead to an over-representation of massive stars in distant clusters, artificially shortening the inferred average disc lifetimes. By mimicking such biases through a stellar mass cut-off in our models, we reproduce both the longer disc lifetimes in nearby clusters ($\gtrsim 5$ Myr) and the shorter ones inferred in more distant systems ($\sim 1$--3 Myr). Furthermore, our findings align with those of \citet{ARibas2014A&A} and \citet{emsenhuber2023towards}, who showed that low-mass stars retain their discs longer when evolving in isolation due to reduced internal photoevaporation. This long-lived disc population may be associated with the so-called {\it Peter Pan} discs \citep{silverberg2020,coleman2022dispersal}, which our model naturally reproduces under conditions of low $\alpha$ and weak external FUV stellar photoevaporation.

 Previously, \citet{Lodato2017MNRAS} presented one of the first comprehensive disc population synthesis  frameworks aimed at interpreting the observed distribution of protoplanetary discs in the disc mass---stellar accretion rate plane. In this work, the authors demonstrated that the observed correlation between disc mass and stellar accretion rate arises naturally from viscous disc evolution when a population of discs with a distribution of initial masses and radii is considered. They also introduced the concept of disc isochrones as a powerful diagnostic tool to interpret disc populations at different evolutionary stages. Our work builds upon this framework but differs from \citet{Lodato2017MNRAS} in several important aspects. While \citet{Lodato2017MNRAS} focused on purely viscous disc evolution and did not include disc dispersal processes, our population synthesis explicitly incorporates both internal and external photoevaporation. This allows us to model disc dispersal 
 and to explore disc lifetimes, accretion rate decay, and disc fractions across different irradiation environments. In this sense, our approach extends the viscous population synthesis framework of \citet{Lodato2017MNRAS} to regimes where mass-loss processes play a dominant role. Another key difference concerns the scientific scope of the two studies. \citet{Lodato2017MNRAS} aimed primarily at explaining the origin of the disc mass---accretion rate correlation at fixed age, without attempting to reproduce observed disc fractions or lifetimes across star-forming regions. In contrast, a central goal of our work is to simultaneously reproduce accretion rates, disc lifetimes, and disc fractions, while accounting for observational biases, time---dependent star formation rates, and environmental effects. This broader approach allows us to place stronger constraints on the range of viable disc parameters and evolutionary pathways.

 More recently, \citet{Somigliana2020, Somigliana2022MNRAS, Somigliana2024} further extended disc population synthesis models by combining viscous evolution with internal photoevaporation and, more recently, MHD winds, with the specific aim of explaining the large observed scatter in the disc mass--stellar accretion rate plane. In particular, \citet{Somigliana2022MNRAS} demonstrated that adopting a stellar-mass-dependent viscous timescale enables their models to reproduce both the observed dispersion in accretion rates and the evolution of the disc mass--accretion rate correlation over time. 

Our work shares several similarities with the approach of \citet{Somigliana2020, Somigliana2022MNRAS, Somigliana2024}, including the use of population synthesis techniques and the recognition that viscous evolution alone is insufficient to capture the full diversity of observed disc properties. In these studies, dispersal mechanisms such as photoevaporation and magnetic winds play a central role in regulating disc evolution. However, there are important differences in the resulting physical interpretation. A key point of divergence concerns the dependence of the viscosity on stellar mass. To match observations, the stellar-mass-dependent timescale in \citet{Somigliana2022MNRAS} implies an average viscosity that decreases with increasing stellar mass, leading to relatively higher viscosities for discs around low-mass stars. In contrast, our results indicate the opposite trend, with $\alpha$ increasing with stellar mass. This difference has direct implications for disc evolutionary timescales. In the framework of \citet{Somigliana2022MNRAS}, enhanced viscosities in low-mass systems lead to more efficient angular momentum transport and therefore to faster disc evolution and dispersal. 
While this assumption successfully reproduces the observed spread in accretion rates, it implies that discs around low-mass stars dissipate more rapidly than those around higher-mass stars. This prediction is in tension with observational evidence indicating that discs around low-mass stars tend to be longer-lived, as inferred from disc fractions and infrared excess studies \citep[e.g.][]{ribas2015protoplanetary,  Downes2015, Pfalzner_2022}. In our model, the increase of $\alpha$ with stellar mass naturally results in longer disc lifetimes for low-mass stars, alleviating this tension with observations. At the same time, we find that the observed dispersion in accretion rates can still be reproduced when the combined effects of viscous evolution, internal photoevaporation, and external irradiation are taken into account, without requiring enhanced viscosities for low-mass systems. This suggests that part of the scatter attributed to viscosity variations in previous studies may instead be explained by environmental effects and multifaceted mass-loss processes.

Another particularly relevant point of comparison is with the study of \citet{emsenhuber2023towards}, who used a similar model to study the disc properties in nearby star-forming regions. Their models indicated that the combined effect of internal and external photoevaporation tends to disperse discs too quickly to match the observed fraction of stars with discs, especially in the clusters of Lupus and Chamaeleon I. To address this, they adopted initial conditions with more massive, compact discs and omitted external photoevaporation entirely. In contrast, our results show that moderate external photoevaporation --when appropriately modelled-- is not only compatible with the observed disc lifetimes but also essential to reproduce low-mass discs with high accretion rates. In our syntheses, external photoevaporation acts preferentially on the outer disc, leaving high surface densities in the inner regions and thus allowing high accretion rates even in low-mass discs --a trend commonly observed in young clusters.

Moreover, \citet{emsenhuber2023towards} adopt a constant $\alpha$ across the stellar mass distribution. This choice fails to reproduce the observed $\dot{M} - M_{\star}$ correlation. In contrast, we introduce a dependence of $\alpha$ on $M_{\star}$, allowing our synthetic populations to recover both the slope and scatter of the observed correlation. We also find that a linearly biased distribution of $\log \alpha_0$ toward higher viscosities further improves agreement with observations.

Another important distinction is our explicit implementation of a time-dependent star formation rate. By incorporating an exponentially declining SFR with a timescale of $t_{\rm d}$ = 1~Myr, we recover the observed increase in disc fractions at early cluster ages. This contrasts with the absence of an SFR assumption in \citet{emsenhuber2023towards}, which limits the model's ability to reproduce age-dependent trends in disc demographics.

Altogether, while both approaches capture aspects of the observed disc population, our study demonstrates that including a stellar-mass-dependent viscosity, extended star formation histories, and moderate external photoevaporation is necessary to simultaneously match the observed diagrams $\dot{M}$--$M_{\rm d}$, $\dot{M}$--$M_{\star}$, and disc lifetimes across various environments.

\subsection{Implications for planet formation}

The diversity of exoplanetary systems observed to date likely reflects the interplay between stellar mass, disc evolution, and environmental effects. Our simulations suggest that different disc dissipation timescales, viscosity regimes, and accretion histories naturally lead to distinct planetary system architectures.

In high-viscosity discs, in our model associated with more massive stars, the combination of rapid viscous evolution and intense internal (and also external) photoevaporation shortens the timescales available for planet formation. On the contrary, these discs are also the more massive ones, favouring for example the formation of gas giants. In fact, occurrence rates of giant planets grow with the stellar mass \citep[see for example][]{Pan2025}. Thus, if giant planets form via core accretion, this implies that massive solid cores should form quickly enough. However, high levels of turbulence can also enhance the collisional fragmentation of small solids and increase their radial diffusion, potentially inhibiting the efficient growth of pebbles and planetesimals. In addition, in the context of pebble accretion, high-viscosity environments reduce its efficiency \citep[e.g][]{JVenturi2020A&Aa}. This effect can act as a bottleneck for the rapid formation of massive cores before the gas disc dissipates, making giant planet formation more difficult unless pebble fluxes are exceptionally high.

Conversely, low-viscosity discs --often found around low-mass stars in our model-- evolve more slowly and are less affected by both internal and external photoevaporation. These discs offer extended lifetimes, which are favourable for the gradual formation of rocky planets, super-Earths, and also possible sub-Neptunes. The reduced turbulence in such environments promotes pebble settling toward the mid-plane, increasing their local density and enhancing pebble accretion efficiency \citep[e.g.][]{JVenturi2020A&Aa}. Furthermore, our results support the idea that external photoevaporation can selectively truncate the outer disc while preserving the high surface density of the inner regions. This configuration can maintain high accretion rates in low-mass discs and may favour the formation of compact, tightly packed planetary systems, similar to those observed in systems like TRAPPIST-1 or Kepler-42.

Taken together, our findings reinforce the notion that the architecture and diversity of planetary systems are deeply influenced by the interplay between stellar mass, disc viscosity, and environmental conditions. Disc population synthesis models that include these dependencies are thus key to bridging the gap between disc observations and the diversity of known exoplanets. 

\section{Conclusions}
\label{sec:conclusions}

One of our main results is that, if discs evolve through viscous accretion and photoevaporation (internal and/or external), it is essential to include a correlation between the disc viscosity and the stellar mass to reproduce the observed correlation between the stellar mass accretion rates and the stellar masses. It remains to be seen whether a physical mechanism with these characteristics can be found. In addition, our synthetic correlation between the stellar mass accretion rates and the stellar masses remains invariant in time, in line with the observational estimations. 

 Furthermore, reproducing the observed high accretion rates requires adopting a probability distribution that favours discs with relatively high viscosities. It is also necessary to account for a finite duration of star formation. In particular, a formation timescale of $\sim 1$ Myr is required to explain the high accretion rates associated with massive stars in young clusters (ages $\sim 1$--$2$ Myr), while extended star formation histories can account for the high accretion rates observed in older regions such as Upper Scorpius. In this context, our population synthesis successfully reproduces the observations of Upper Scorpius assuming a formation timescale of $\sim 5$ Myr.

 Including a distribution of formation times naturally explains the observed dispersion in accretion rates and leads to the presence of high accretion rates in moderate- and high-mass discs. However, in both young clusters ($\sim 1$--$2$ Myr) and older regions, reproducing the population of discs with high accretion rates and low inferred masses requires invoking either external photoevaporation or a systematic underestimation of the gas disc mass due to dust evolution processes, including grain growth, radial drift, dust trapping, and planetesimal formation.

Finally, we find that our synthetic population in which disc viscosities are correlated with stellar masses is able to reproduce the estimated ages of protoplanetary discs --ranging from 1 to 20 Myr--, and the fraction of stars with discs observed in young stellar clusters. In addition, setting a lower limit for the stellar mass of our population (mimicking the distances to the different young stellar clusters and the observational star mass limit) we are also able to reproduce the different observational estimations of the disc fractions for young stellar clusters at different distances.  

\begin{acknowledgements}
We thank Alexander Emsenhuber, Remo Burn and Jesse Weder for useful discussions during the development of this work. We would like to thank the anonymous referee for his/her rigorous review and valuable comments, which helped us to improve the manuscript. This work is partially supported by PIP 2971 from CONICET (Argentina) and by PID-SG004 from UNLP (Argentina). We also thank Juan Ignacio Rodriguez from IALP for the computation managing resources of the Grupo de Astrof\'{\i}sica Planetaria de La Plata.
\end{acknowledgements}

\bibliographystyle{aa} 
\bibliography{bibliografia} 

@ARTICLE{anania2025a,
       author = {{Anania}, Rossella and {Rosotti}, Giovanni P. and {G{\'a}rate}, Mat{\'\i}as and {Pinilla}, Paola and {Vioque}, Miguel and {Trapman}, Leon and {Carpenter}, John and {Zhang}, Ke and {Pascucci}, Ilaria and {Cieza}, Lucas A. and {Sierra}, Anibal and {Kurtovic}, Nicolas T. and {Miley}, James and {P{\'e}rez}, Laura M. and {Tabone}, Beno{\^\i}t and {Hogerheijde}, Michiel and {Deng}, Dingshan and {Agurto-Gangas}, Carolina and {Ruiz-Rodriguez}, Dary A. and {Gonz{\'a}lez-Ruilova}, Camilo and {TorresVillanueva}, Estephani E.},
        title = "{The ALMA Survey of Gas Evolution of PROtoplanetary Disks (AGE-PRO). VIII. The Impact of External Photoevaporation on Disk Masses and Radii in Upper Scorpius}",
      journal = {\apj},
         year = 2025,
        month = aug,
       volume = {989},
       number = {1},
          eid = {8},
        pages = {8},
          doi = {10.3847/1538-4357/adb587},
archivePrefix = {arXiv},
       eprint = {2506.10743},
 primaryClass = {astro-ph.EP},
       adsurl = {https://ui.adsabs.harvard.edu/abs/2025ApJ...989....8A}
}

@ARTICLE{longarini2025,
       author = {{Longarini}, Cristiano and {Lodato}, Giuseppe and {Rosotti}, Giovanni and {Andrews}, Sean and {Winter}, Andrew and {Stadler}, Jochen and {Izquierdo}, Andr{\'e}s and {Galloway-Sprietsma}, Maria and {Facchini}, Stefano and {Curone}, Pietro and {Benisty}, Myriam and {Teague}, Richard and {Bae}, Jaehan and {Barraza-Alfaro}, Marcelo and {Cataldi}, Gianni and {Czekala}, Ian and {Cuello}, Nicol{\'a}s and {Fasano}, Daniele and {Flock}, Mario and {Fukagawa}, Misato and {Garg}, Himanshi and {Hall}, Cassandra and {Hammond}, Iain and {Hardiman}, Caitlyn and {Hilder}, Thomas and {Huang}, Jane and {Ilee}, John D. and {Isella}, Andrea and {Kanagawa}, Kazuhiro and {Lesur}, Geoffroy and {Loomis}, Ryan A. and {M{\'e}nard}, Francois and {Orihara}, Ryuta and {Pinte}, Christophe and {Price}, Daniel and {Testi}, Leonardo and {Fernandez}, Gaylor Wafflard- and {W{\"o}lfer}, Lisa and {Yen}, Hsi-Wei and {Yoshida}, Tomohiro C. and {Zawadzki}, Brianna},
        title = "{exoALMA. XII. Weighing and Sizing exoALMA Disks with Rotation Curve Modelling}",
      journal = {\apjl},
         year = 2025,
        month = may,
       volume = {984},
       number = {1},
          eid = {L17},
        pages = {L17},
          doi = {10.3847/2041-8213/adc431},
archivePrefix = {arXiv},
       eprint = {2504.18726},
 primaryClass = {astro-ph.EP},
       adsurl = {https://ui.adsabs.harvard.edu/abs/2025ApJ...984L..17L}
}

@ARTICLE{savvidou2025,
       author = {{Savvidou}, Sofia and {Bitsch}, Bertram},
        title = "{There is no disk mass budget problem of planet formation}",
      journal = {\aap},
         year = 2025,
        month = jan,
       volume = {693},
          eid = {A302},
        pages = {A302},
          doi = {10.1051/0004-6361/202449847},
archivePrefix = {arXiv},
       eprint = {2407.08533},
 primaryClass = {astro-ph.EP},
       adsurl = {https://ui.adsabs.harvard.edu/abs/2025A&A...693A.302S}
}

@ARTICLE{Somigliana2024,
       author = {{Somigliana}, Alice and {Testi}, Leonardo and {Rosotti}, Giovanni and {Toci}, Claudia and {Lodato}, Giuseppe and {Anania}, Rossella and {Tabone}, Beno{\^\i}t and {Tazzari}, Marco and {Klessen}, Ralf and {Lebreuilly}, Ugo and {Hennebelle}, Patrick and {Molinari}, Sergo},
        title = "{The evolution of the M$_{d}${\textendash}M$_{{\ensuremath{\star}}}$ and Ṁ{\textendash}M$_{{\ensuremath{\star}}}$ correlations traces protoplanetary disc dispersal}",
      journal = {\aap},
         year = 2024,
        month = sep,
       volume = {689},
          eid = {A285},
        pages = {A285},
          doi = {10.1051/0004-6361/202450744},
archivePrefix = {arXiv},
       eprint = {2407.21101},
 primaryClass = {astro-ph.EP},
       adsurl = {https://ui.adsabs.harvard.edu/abs/2024A&A...689A.285S}
}

@ARTICLE{Ronco2024,
       author = {{Ronco}, M.~P. and {Schreiber}, M.~R. and {Villaver}, E. and {Guilera}, O.~M. and {Miller Bertolami}, M.~M.},
        title = "{Planet formation around intermediate-mass stars. I. Different disc evolutionary pathways as a function of stellar mass}",
      journal = {\aap},
         year = 2024,
        month = feb,
       volume = {682},
          eid = {A155},
        pages = {A155},
          doi = {10.1051/0004-6361/202347762},
archivePrefix = {arXiv},
       eprint = {2311.03934},
 primaryClass = {astro-ph.EP},
       adsurl = {https://ui.adsabs.harvard.edu/abs/2024A&A...682A.155R}
}

@ARTICLE{2024ApJPfalzner,
       author = {{Pfalzner}, Susanne and {Dincer}, Furkan},
        title = "{Low-mass Stars: Their Protoplanetary Disk Lifetime Distribution}",
      journal = {\apj},
         year = 2024,
        month = mar,
       volume = {963},
       number = {2},
          eid = {122},
        pages = {122},
          doi = {10.3847/1538-4357/ad1bef},
archivePrefix = {arXiv},
       eprint = {2401.03775},
 primaryClass = {astro-ph.EP},
       adsurl = {https://ui.adsabs.harvard.edu/abs/2024ApJ...963..122P}
}

@ARTICLE{2023Rosotti,
       author = {{Rosotti}, Giovanni P.},
        title = "{Empirical constraints on turbulence in proto-planetary discs}",
      journal = {\nar},
         year = 2023,
        month = jun,
       volume = {96},
          eid = {101674},
        pages = {101674},
          doi = {10.1016/j.newar.2023.101674},
archivePrefix = {arXiv},
       eprint = {2302.01433},
 primaryClass = {astro-ph.EP},
       adsurl = {https://ui.adsabs.harvard.edu/abs/2023NewAR..9601674R}
}

@ARTICLE{emsenhuber2023towards,
       author = {{Emsenhuber}, Alexandre and {Burn}, Remo and {Weder}, Jesse and {Monsch}, Kristina and {Picogna}, Giovanni and {Ercolano}, Barbara and {Preibisch}, Thomas},
        title = "{Toward a population synthesis of disks and planets. II. Confronting disk models and observations at the population level}",
      journal = {\aap},
         year = 2023,
        month = may,
       volume = {673},
          eid = {A78},
        pages = {A78},
          doi = {10.1051/0004-6361/202244767},
archivePrefix = {arXiv},
       eprint = {2301.04656},
 primaryClass = {astro-ph.EP},
       adsurl = {https://ui.adsabs.harvard.edu/abs/2023A&A...673A..78E}
}

@INPROCEEDINGS{manara2023ASPC,
       author = {{Manara}, C.~F. and {Ansdell}, M. and {Rosotti}, G.~P. and {Hughes}, A.~M. and {Armitage}, P.~J. and {Lodato}, G. and {Williams}, J.~P.},
        title = "{Demographics of Young Stars and their Protoplanetary Disks: Lessons Learned on Disk Evolution and its Connection to Planet Formation}",
    booktitle = {Protostars and Planets VII},
         year = 2023,
       editor = {{Inutsuka}, S. and {Aikawa}, Y. and {Muto}, T. and {Tomida}, K. and {Tamura}, M.},
       series = {Astronomical Society of the Pacific Conference Series},
       volume = {534},
        month = jul,
        pages = {539},
          doi = {10.48550/arXiv.2203.09930},
archivePrefix = {arXiv},
       eprint = {2203.09930},
 primaryClass = {astro-ph.SR},
       adsurl = {https://ui.adsabs.harvard.edu/abs/2023ASPC..534..539M}
}

@ARTICLE{Somigliana2022MNRAS,
       author = {{Somigliana}, Alice and {Toci}, Claudia and {Rosotti}, Giovanni and {Lodato}, Giuseppe and {Tazzari}, Marco and {Manara}, Carlo F. and {Testi}, Leonardo and {Lepri}, Federico},
        title = "{On the time evolution of the M$_{d}$-M$_{{\ensuremath{\star}}}$ and Ṁ-M$_{{\ensuremath{\star}}}$ correlations for protoplanetary discs: the viscous time-scale increases with stellar mass}",
      journal = {\mnras},
         year = 2022,
        month = aug,
       volume = {514},
       number = {4},
        pages = {5927-5940},
          doi = {10.1093/mnras/stac1587},
archivePrefix = {arXiv},
       eprint = {2206.04136},
 primaryClass = {astro-ph.EP},
       adsurl = {https://ui.adsabs.harvard.edu/abs/2022MNRAS.514.5927S}
}

@ARTICLE{haworth2018fried,
       author = {{Haworth}, Thomas J. and {Clarke}, Cathie J. and {Rahman}, Wahidur and {Winter}, Andrew J. and {Facchini}, Stefano},
        title = "{The FRIED grid of mass-loss rates for externally irradiated protoplanetary discs}",
      journal = {\mnras},
         year = 2018,
        month = nov,
       volume = {481},
       number = {1},
        pages = {452-466},
          doi = {10.1093/mnras/sty2323},
archivePrefix = {arXiv},
       eprint = {1808.07484},
 primaryClass = {astro-ph.SR},
       adsurl = {https://ui.adsabs.harvard.edu/abs/2018MNRAS.481..452H}
}

@article{Pfalzner_2022,
doi = {10.3847/2041-8213/ac9839},
url = {https://dx.doi.org/10.3847/2041-8213/ac9839},
year = {2022},
month = {oct},
publisher = {The American Astronomical Society},
volume = {939},
number = {1},
pages = {L10},
author = {Susanne Pfalzner and Shahrzad Dehghani and Arnaud Michel},
title = {Most Planets Might Have More than 5 Myr of Time to Form},
journal = {\apjl}
}

@article{coleman2022dispersal,
      title={Dispersal of protoplanetary discs: How stellar properties and the local environment determine the pathway of evolution},
      author={Coleman, Gavin AL and Haworth, Thomas J},
      journal={\mnras},
      year={2022}
    }

@article{drazkowska2022planet,
      title={Planet Formation Theory in the Era of ALMA and Kepler: from Pebbles to Exoplanets},
      author={Drazkowska, Joanna and Bitsch, Bertram and Lambrechts, Michiel and Mulders, Gijs D and Harsono, Daniel and Vazan, Allona and Liu, Beibei and Ormel, Chris W and Kretke, Katherine and Morbidelli, Alessandro},
      journal={Protostars and Planets VII, ASP Electronic Monographs},
      pages={717},
      year={2023}
}

@ARTICLE{2021Squicciarini,
       author = {{Squicciarini}, Vito and {Gratton}, Raffaele and {Bonavita}, Mariangela and {Mesa}, Dino},
        title = "{Unveiling the star formation history of the Upper Scorpius association through its kinematics}",
      journal = {\mnras},
         year = 2021,
        month = oct,
       volume = {507},
       number = {1},
        pages = {1381-1400},
          doi = {10.1093/mnras/stab2079},
archivePrefix = {arXiv},
       eprint = {2107.08057},
 primaryClass = {astro-ph.GA},
       adsurl = {https://ui.adsabs.harvard.edu/abs/2021MNRAS.507.1381S}
}

@ARTICLE{Kunitomo2021,
       author = {{Kunitomo}, Masanobu and {Ida}, Shigeru and {Takeuchi}, Taku and {Pani{\'c}}, Olja and {Miley}, James M. and {Suzuki}, Takeru K.},
        title = "{Photoevaporative Dispersal of Protoplanetary Disks around Evolving Intermediate-mass Stars}",
      journal = {\apj},
         year = 2021,
        month = mar,
       volume = {909},
       number = {2},
          eid = {109},
        pages = {109},
          doi = {10.3847/1538-4357/abdb2a},
archivePrefix = {arXiv},
       eprint = {2103.07673},
 primaryClass = {astro-ph.EP},
       adsurl = {https://ui.adsabs.harvard.edu/abs/2021ApJ...909..109K}
}

@ARTICLE{Emsenhuber_2021,
       author = {{Emsenhuber}, Alexandre and {Mordasini}, Christoph and {Burn}, Remo and {Alibert}, Yann and {Benz}, Willy and {Asphaug}, Erik},
        title = "{The New Generation Planetary Population Synthesis (NGPPS). II. Planetary population of solar-like stars and overview of statistical results}",
      journal = {\aap},
         year = 2021,
        month = dec,
       volume = {656},
          eid = {A70},
        pages = {A70},
          doi = {10.1051/0004-6361/202038863},
archivePrefix = {arXiv},
       eprint = {2007.05562},
 primaryClass = {astro-ph.EP},
       adsurl = {https://ui.adsabs.harvard.edu/abs/2021A&A...656A..70E}
}

@ARTICLE{Somigliana2020,
       author = {{Somigliana}, Alice and {Toci}, Claudia and {Lodato}, Giuseppe and {Rosotti}, Giovanni and {Manara}, Carlo F.},
        title = "{Effects of photoevaporation on protoplanetary disc `isochrones'}",
      journal = {\mnras},
         year = 2020,
        month = feb,
       volume = {492},
       number = {1},
        pages = {1120-1126},
          doi = {10.1093/mnras/stz3481},
archivePrefix = {arXiv},
       eprint = {1912.05623},
 primaryClass = {astro-ph.EP},
       adsurl = {https://ui.adsabs.harvard.edu/abs/2020MNRAS.492.1120S}
}

@ARTICLE{Venturini20Review,
       author = {{Venturini}, Julia and {Ronco}, Maria Paula and {Guilera}, Octavio Miguel},
        title = "{Setting the Stage: Planet Formation and Volatile Delivery}",
      journal = {\ssr},
         year = {2020},
        month = jul,
       volume = {216},
       number = {5},
          eid = {86},
        pages = {86},
          doi = {10.1007/s11214-020-00700-y},
archivePrefix = {arXiv},
       eprint = {2006.07127},
 primaryClass = {astro-ph.EP},
       adsurl = {https://ui.adsabs.harvard.edu/abs/2020SSRv..216...86V}
}

@ARTICLE{Andrews2020ARAA,
       author = {{Andrews}, Sean M.},
        title = "{Observations of Protoplanetary Disk Structures}",
      journal = {\araa},
         year = {2020},
        month = aug,
       volume = {58},
        pages = {483-528},
          doi = {10.1146/annurev-astro-031220-010302},
archivePrefix = {arXiv},
       eprint = {2001.05007},
 primaryClass = {astro-ph.EP},
       adsurl = {https://ui.adsabs.harvard.edu/abs/2020ARA&A..58..483A}
}

@ARTICLE{JVenturi2020A&Aa,
       author = {{Venturini}, Julia and {Guilera}, Octavio Miguel and {Ronco}, Mar{\'\i}a Paula and {Mordasini}, Christoph},
        title = "{Most super-Earths formed by dry pebble accretion are less massive than 5 Earth masses}",
      journal = {\aap},
         year = 2020,
        month = dec,
       volume = {644},
          eid = {A174},
        pages = {A174},
          doi = {10.1051/0004-6361/202039140},
archivePrefix = {arXiv},
       eprint = {2008.05497},
 primaryClass = {astro-ph.EP},
       adsurl = {https://ui.adsabs.harvard.edu/abs/2020A&A...644A.174V}
}

@ARTICLE{ Garate2020A&A,
       author = {{G{\'a}rate}, Mat{\'\i}as and {Birnstiel}, Til and {Dr{\k{a}}{\.z}kowska}, Joanna and {Stammler}, Sebastian Markus},
        title = "{Gas accretion damped by dust back-reaction at the snow line}",
      journal = {\aap},
         year = 2020,
        month = mar,
       volume = {635},
          eid = {A149},
        pages = {A149},
          doi = {10.1051/0004-6361/201936067},
archivePrefix = {arXiv},
       eprint = {1906.07708},
 primaryClass = {astro-ph.EP},
       adsurl = {https://ui.adsabs.harvard.edu/abs/2020A&A...635A.149G}
}

@article{Rosotti2019MN,
    author = {Rosotti, Giovanni P and Tazzari, Marco and Booth, Richard A and Testi, Leonardo and Lodato, Giuseppe and Clarke, Cathie},
    title = {The time evolution of dusty protoplanetary disc radii: observed and physical radii differ},
    journal = {\mnras},
    volume = {486},
    number = {4},
    pages = {4829-4844},
    year = {2019},
    month = {05},
    issn = {0035-8711},
    doi = {10.1093/mnras/stz1190},
    url = {https://doi.org/10.1093/mnras/stz1190},
    eprint = {https://academic.oup.com/mnras/article-pdf/486/4/4829/28599251/stz1190.pdf},
}

@ARTICLE{2019A&A.Manara,
       author = {{Manara}, C.~F. and {Mordasini}, C. and {Testi}, L. and {Williams}, J.~P. and {Miotello}, A. and {Lodato}, G. and {Emsenhuber}, A.},
        title = "{Constraining disk evolution prescriptions of planet population synthesis models with observed disk masses and accretion rates}",
      journal = {\aap},
         year = 2019,
        month = nov,
       volume = {631},
          eid = {L2},
        pages = {L2},
          doi = {10.1051/0004-6361/201936488},
archivePrefix = {arXiv},
       eprint = {1909.08485},
 primaryClass = {astro-ph.EP},
       adsurl = {https://ui.adsabs.harvard.edu/abs/2019A&A...631L...2M}
}

@ARTICLE{OGuilera2019MNRASb,
       author = {{Guilera}, O.~M. and {Cuello}, N. and {Montesinos}, M. and {Miller Bertolami}, M.~M. and {Ronco}, M.~P. and {Cuadra}, J. and {Masset}, F.~S.},
        title = "{Thermal torque effects on the migration of growing low-mass planets}",
      journal = {\mnras},
         year = 2019,
        month = jul,
       volume = {486},
       number = {4},
        pages = {5690-5708},
          doi = {10.1093/mnras/stz1158},
archivePrefix = {arXiv},
       eprint = {1904.11047},
 primaryClass = {astro-ph.EP},
       adsurl = {https://ui.adsabs.harvard.edu/abs/2019MNRAS.486.5690G}
}

@article{Sellek2019,
    author = {Sellek, Andrew D and Booth, Richard A and Clarke, Cathie J},
    title = "{The evolution of dust in discs influenced by external photoevaporation}",
    journal = {\mnras},
    volume = {492},
    number = {1},
    pages = {1279-1294},
    year = {2019},
    month = {12},
    issn = {0035-8711},
    doi = {10.1093/mnras/stz3528},
    url = {https://doi.org/10.1093/mnras/stz3528},
    eprint = {https://academic.oup.com/mnras/article-pdf/492/1/1279/31784642/stz3528.pdf},
}

@article{Flaherty_2018,
doi = {10.3847/1538-4357/aab615},
url = {https://dx.doi.org/10.3847/1538-4357/aab615},
year = {2018},
month = {mar},
publisher = {The American Astronomical Society},
volume = {856},
number = {2},
pages = {117},
author = {Kevin M. Flaherty and A. Meredith Hughes and Richard Teague and Jacob B. Simon and Sean M. Andrews and David J. Wilner},
title = {Turbulence in the TW Hya Disk},
journal = {\apj}
}

@article{tychoniec2018vla,
  title={The VLA Nascent Disk and Multiplicity Survey of Perseus Protostars (VANDAM). IV. Free--Free Emission from Protostars: Links to infrared properties, outflow tracers, and protostellar disk masses},
  author={Tychoniec, {\L}ukasz and Tobin, John J and Karska, Agata and Chandler, Claire and Dunham, Michael M and Harris, Robert J and Kratter, Kaitlin M and Li, Zhi-Yun and Looney, Leslie W and Melis, Carl and others},
  journal={\apjs},
  volume={238},
  number={2},
  pages={19},
  year={2018},
  publisher={IOP Publishing}
}

@ARTICLE{Lodato2017MNRAS,
       author = {{Lodato}, Giuseppe and {Scardoni}, Chiara E. and {Manara}, Carlo F. and {Testi}, Leonardo},
        title = "{Protoplanetary disc `isochrones' and the evolution of discs in the M˙-M$_{d}$ plane}",
      journal = {\mnras},
         year = 2017,
        month = dec,
       volume = {472},
       number = {4},
        pages = {4700-4706},
          doi = {10.1093/mnras/stx2273},
archivePrefix = {arXiv},
       eprint = {1708.09467},
 primaryClass = {astro-ph.SR},
       adsurl = {https://ui.adsabs.harvard.edu/abs/2017MNRAS.472.4700L}
}

@ARTICLE{2017A&ATazzari,
       author = {{Tazzari}, M. and {Testi}, L. and {Natta}, A. and {Ansdell}, M. and {Carpenter}, J. and {Guidi}, G. and {Hogerheijde}, M. and {Manara}, C.~F. and {Miotello}, A. and {van der Marel}, N. and {van Dishoeck}, E.~F. and {Williams}, J.~P.},
        title = "{Physical properties of dusty protoplanetary disks in Lupus: evidence for viscous evolution?}",
      journal = {\aap},
         year = 2017,
        month = oct,
       volume = {606},
          eid = {A88},
        pages = {A88},
          doi = {10.1051/0004-6361/201730890},
archivePrefix = {arXiv},
       eprint = {1707.01499},
 primaryClass = {astro-ph.EP},
       adsurl = {https://ui.adsabs.harvard.edu/abs/2017A&A...606A..88T}
}

@article{pronco2017,
    author = {Ronco, M. P. and Guilera, O. M. and de Elía, G. C.},
    title = "{Formation of solar system analogues – I. Looking for initial conditions through a population synthesis analysis}",
    journal = {\mnras},
    volume = {471},
    number = {3},
    pages = {2753-2770},
    year = {2017},
    month = {07},
    issn = {0035-8711},
    doi = {10.1093/mnras/stx1746},
    url = {https://doi.org/10.1093/mnras/stx1746},
    eprint = {https://academic.oup.com/mnras/article-pdf/471/3/2753/19496935/stx1746.pdf},
}

@PHDTHESIS{Ansdell017,
       author = {{Ansdell}, Megan C.},
        title = "{Protoplanetary disk demographics with ALMA}",
       school = {University of Hawaii, Manoa},
         year = 2017,
        month = jan,
       adsurl = {https://ui.adsabs.harvard.edu/abs/2017PhDT.......196A}
}

@ARTICLE{2017Ecol,
       author = {{Ercolano}, Barbara and {Pascucci}, Ilaria},
        title = "{The dispersal of planet-forming discs: theory confronts observations}",
      journal = {Royal Society Open Science},
         year = 2017,
        month = apr,
       volume = {4},
       number = {4},
          eid = {170114},
        pages = {170114},
          doi = {10.1098/rsos.170114},
archivePrefix = {arXiv},
       eprint = {1704.00214},
 primaryClass = {astro-ph.EP},
       adsurl = {https://ui.adsabs.harvard.edu/abs/2017RSOS....470114E}
}

@article{OctavioMarcelo2017Codigo,
    author = {Guilera, O. M. and Miller Bertolami, M. M. and Ronco, M. P.},
    title = "{The formation of giant planets in wide orbits by photoevaporation-synchronized migration}",
    journal = {\mnras},
    volume = {471},
    number = {1},
    pages = {L16-L20},
    year = {2017},
    month = {06},
    issn = {1745-3925},
    doi = {10.1093/mnrasl/slx095},
    url = {https://doi.org/10.1093/mnrasl/slx095},
    eprint = {https://academic.oup.com/mnrasl/article-pdf/471/1/L16/18562741/slx095.pdf},
}

@ARTICLE{2016A&AManara,
       author = {{Manara}, C.~F. and {Rosotti}, G. and {Testi}, L. and {Natta}, A. and {Alcal{\'a}}, J.~M. and {Williams}, J.~P. and {Ansdell}, M. and {Miotello}, A. and {van der Marel}, N. and {Tazzari}, M. and {Carpenter}, J. and {Guidi}, G. and {Mathews}, G.~S. and {Oliveira}, I. and {Prusti}, T. and {van Dishoeck}, E.~F.},
        title = "{Evidence for a correlation between mass accretion rates onto young stars and the mass of their protoplanetary disks}",
      journal = {\aap},
         year = 2016,
        month = jun,
       volume = {591},
          eid = {L3},
        pages = {L3},
          doi = {10.1051/0004-6361/201628549},
archivePrefix = {arXiv},
       eprint = {1605.03050},
 primaryClass = {astro-ph.SR},
       adsurl = {https://ui.adsabs.harvard.edu/abs/2016A&A...591L...3M}
}

@article{ribas2015protoplanetary,
  title={Protoplanetary disk lifetimes vs. stellar mass and possible implications for giant planet populations},
  author={Ribas, {\'A}lvaro and Bouy, Herv{\'e} and Mer{\'\i}n, Bruno},
  journal={\aap},
  volume={576},
  pages={A52},
  year={2015},
  publisher={EDP Sciences}
}

@ARTICLE{ARibas2014A&A,
       author = {{Ribas}, {\'A}lvaro and {Mer{\'\i}n}, Bruno and {Bouy}, Herv{\'e} and {Maud}, Luke T.},
        title = "{Disk evolution in the solar neighborhood. I. Disk frequencies from 1 to 100 Myr}",
      journal = {\aap},
         year = 2014,
        month = jan,
       volume = {561},
          eid = {A54},
        pages = {A54},
          doi = {10.1051/0004-6361/201322597},
archivePrefix = {arXiv},
       eprint = {1312.0609},
 primaryClass = {astro-ph.SR},
       adsurl = {https://ui.adsabs.harvard.edu/abs/2014A&A...561A..54R}
}

@article{Pfalzner_2014,
    doi = {10.1088/2041-8205/793/2/L34},
    url = {https://dx.doi.org/10.1088/2041-8205/793/2/L34},
    year = {2014},
    month = {sep},
    publisher = {The American Astronomical Society},
    volume = {793},
    number = {2},
    pages = {L34},
    author = {Susanne Pfalzner and Manuel Steinhausen and Karl Menten},
    title = {SHORT DISSIPATION TIMES OF PROTO-PLANETARY DISKS: AN ARTIFACT OF SELECTION EFFECTS?},
    journal = {\apjl}
    }

@article{owen2012theoryfoto,
  title={On the theory of disc photoevaporation},
  author={Owen, James E and Clarke, Cathie J and Ercolano, Barbara},
  journal={\mnras},
  volume={422},
  number={3},
  pages={1880--1901},
  year={2012},
  publisher={The Royal Astronomical Society}
}

@article{Andrews_2010,
	doi = {10.1088/0004-637x/723/2/1241},
	url = {https://doi.org/10.1088/2F0004-637x/2F723/2F2/2F1241},
	year = {2010},
	month = {oct},
	publisher = {American Astronomical Society},
	volume = {723},
	number = {2},
	pages = {1241--1254},
	author = {Sean M. Andrews and D. J. Wilner and A. M. Hughes and Chunhua Qi and C. P. Dullemond},
	title = {{PROTOPLANETARY} {DISK} {STRUCTURES} {IN} {OPHIUCHUS}. {II}. {EXTENSION} {TO} {FAINTER} {SOURCES}},
	journal = {\apj}
}

@ARTICLE{owen2011,
       author = {{Owen}, James E. and {Ercolano}, Barbara and {Clarke}, Cathie J.},
        title = "{Protoplanetary disc evolution and dispersal: the implications of X-ray photoevaporation}",
      journal = {\mnras},
         year = 2011,
        month = mar,
       volume = {412},
       number = {1},
        pages = {13-25},
          doi = {10.1111/j.1365-2966.2010.17818.x},
archivePrefix = {arXiv},
       eprint = {1010.0826},
 primaryClass = {astro-ph.SR},
       adsurl = {https://ui.adsabs.harvard.edu/abs/2011MNRAS.412...13O}
}

@INPROCEEDINGS{MamajekE,
       author = {{Mamajek}, Eric E.},
        title = "{Initial Conditions of Planet Formation: Lifetimes of Primordial Disks}",
    booktitle = {Exoplanets and Disks: Their Formation and Diversity},
         year = 2009,
       editor = {{Usuda}, Tomonori and {Tamura}, Motohide and {Ishii}, Miki},
       series = {American Institute of Physics Conference Series},
       volume = {1158},
        month = aug,
    publisher = {AIP},
        pages = {3-10},
          doi = {10.1063/1.3215910},
archivePrefix = {arXiv},
       eprint = {0906.5011},
 primaryClass = {astro-ph.EP},
       adsurl = {https://ui.adsabs.harvard.edu/abs/2009AIPC.1158....3M}
}

@inbook{armitage_2009_evolution, place={Cambridge}, title={Protoplanetary disk evolution}, DOI={10.1017/CBO9780511802225.004}, booktitle={Astrophysics of Planet Formation}, publisher={Cambridge University Press}, author={Armitage, Philip J.}, year={2009}, pages={65–108}}

@article{Gorti_2009,
doi = {10.1088/0004-637X/705/2/1237},
url = {https://dx.doi.org/10.1088/0004-637X/705/2/1237},
year = {2009},
month = {oct},
publisher = {The American Astronomical Society},
volume = {705},
number = {2},
pages = {1237},
author = {U. Gorti and C. P. Dullemond and D. Hollenbach},
title = {TIME EVOLUTION OF VISCOUS CIRCUMSTELLAR DISKS DUE TO PHOTOEVAPORATION BY FAR-ULTRAVIOLET, EXTREME-ULTRAVIOLET, AND X-RAY RADIATION FROM THE CENTRAL STAR},
journal = {\apj}
}

@ARTICLE{Suzuki2009,
       author = {{Suzuki}, Takeru K. and {Inutsuka}, Shu-ichiro},
        title = "{Disk Winds Driven by Magnetorotational Instability and Dispersal of Protoplanetary Disks}",
      journal = {\apjl},
         year = 2009,
        month = jan,
       volume = {691},
       number = {1},
        pages = {L49-L54},
          doi = {10.1088/0004-637X/691/1/L49},
archivePrefix = {arXiv},
       eprint = {0812.0844},
 primaryClass = {astro-ph},
       adsurl = {https://ui.adsabs.harvard.edu/abs/2009ApJ...691L..49S}
}

@article{clarke2001,
    author = {Clarke, C. J. and Gendrin, A. and Sotomayor, M.},
    title = {The dispersal of circumstellar discs: the role of the ultraviolet switch},
    journal = {\mnras},
    volume = {328},
    number = {2},
    pages = {485-491},
    year = {2001},
    month = {12},
    issn = {0035-8711},
    doi = {10.1046/j.1365-8711.2001.04891.x},
    url = {https://doi.org/10.1046/j.1365-8711.2001.04891.x},
    eprint = {https://academic.oup.com/mnras/article-pdf/328/2/485/3791576/328-2-485.pdf},
}

@article{kroupa2001,
    author = {Kroupa, Pavel},
    title = "{On the variation of the initial mass function}",
    journal = {\mnras},
    volume = {322},
    number = {2},
    pages = {231-246},
    year = {2001},
    month = {04},
    issn = {0035-8711},
    doi = {10.1046/j.1365-8711.2001.04022.x},
    url = {https://doi.org/10.1046/j.1365-8711.2001.04022.x},
    eprint = {https://academic.oup.com/mnras/article-pdf/322/2/231/2852412/322-2-231.pdf},
}

@article{Hartmann_1998,
doi = {10.1086/305277},
url = {https://dx.doi.org/10.1086/305277},
year = {1998},
month = {mar},
publisher = {},
volume = {495},
number = {1},
pages = {385},
author = {Lee Hartmann and Nuria Calvet and Erik Gullbring and Paola D'Alessio},
title = {Accretion and the Evolution of T Tauri Disks},
journal = {\apj}
}

@article{pringle1981accretion,
  title={Accretion discs in astrophysics},
  author={Pringle, James E},
  journal={Annual review of astronomy and astrophysics},
  volume={19},
  pages={137--162},
  year={1981}
}

@ARTICLE{BellPringle1974,
       author = {{Lynden-Bell}, D. and {Pringle}, J.~E.},
        title = "{The evolution of viscous discs and the origin of the nebular variables.}",
      journal = {\mnras},
         year = {1974},
        month = sep,
       volume = {168},
        pages = {603-637},
          doi = {10.1093/mnras/168.3.603},
       adsurl = {https://ui.adsabs.harvard.edu/abs/1974MNRAS.168..603L}
}

@article{shakura1973black,
  title={Black holes in binary systems. Observational appearance.},
  author={Shakura, Ni I and Sunyaev, Rashid Alievich},
  journal={\aap},
  volume={24},
  pages={337--355},
  year={1973}
}

@ARTICLE{Toomre1964ApJ...139.1217T,
       author = {{Toomre}, A.},
        title = "{On the gravitational stability of a disk of stars.}",
      journal = {\apj},
         year = {1964},
        month = may,
       volume = {139},
        pages = {1217-1238},
          doi = {10.1086/147861},
       adsurl = {https://ui.adsabs.harvard.edu/abs/1964ApJ...139.1217T}
}

@ARTICLE{fang2017,
       author = {{Fang}, Qiliang and {Herczeg}, Gregory J. and {Rizzuto}, Aaron},
        title = "{Age Spreads and the Temperature Dependence of Age Estimates in Upper Sco}",
      journal = {\apj},
         year = 2017,
        month = jun,
       volume = {842},
       number = {2},
          eid = {123},
        pages = {123},
          doi = {10.3847/1538-4357/aa74ca},
archivePrefix = {arXiv},
       eprint = {1705.08612},
 primaryClass = {astro-ph.SR},
       adsurl = {https://ui.adsabs.harvard.edu/abs/2017ApJ...842..123F}
}

@ARTICLE{Pan2025,
       author = {{Pan}, Mengrui and {Liu}, Beibei and {Jiang}, Linjie and {Xie}, Jiwei and {Zhu}, Wei and {Ribas}, Ignasi},
        title = "{Dependence of Planet Populations on Stellar Mass and Metallicity: A Pebble-accretion-based Planet Population Synthesis Model}",
      journal = {\apj},
         year = 2025,
        month = may,
       volume = {985},
       number = {1},
          eid = {7},
        pages = {7},
          doi = {10.3847/1538-4357/adc7a9},
archivePrefix = {arXiv},
       eprint = {2504.00296},
 primaryClass = {astro-ph.EP},
       adsurl = {https://ui.adsabs.harvard.edu/abs/2025ApJ...985....7P}
}

@ARTICLE{Venturini+2024,
       author = {{Venturini}, J. and {Ronco}, M.~P. and {Guilera}, O.~M. and {Haldemann}, J. and {Mordasini}, C. and {Miller Bertolami}, M.},
        title = "{A fading radius valley towards M dwarfs, a persistent density valley across stellar types}",
      journal = {\aap},
         year = 2024,
        month = jun,
       volume = {686},
          eid = {L9},
        pages = {L9},
          doi = {10.1051/0004-6361/202349088},
archivePrefix = {arXiv},
       eprint = {2404.01967},
 primaryClass = {astro-ph.EP},
       adsurl = {https://ui.adsabs.harvard.edu/abs/2024A&A...686L...9V}
}

@ARTICLE{blandford1982,
       author = {{Blandford}, R.~D. and {Payne}, D.~G.},
        title = "{Hydromagnetic flows from accretion disks and the production of radio jets.}",
      journal = {\mnras},
         year = 1982,
        month = jun,
       volume = {199},
        pages = {883-903},
          doi = {10.1093/mnras/199.4.883},
       adsurl = {https://ui.adsabs.harvard.edu/abs/1982MNRAS.199..883B}
}

@article{delfini2025,
    title = {Star formation and accretion rates within 500 pc as traced by {Gaia} {DR3} {XP} spectra},
    volume = {699},
    issn = {0004-6361},
    url = {https://ui.adsabs.harvard.edu/abs/2025A&A...699A.145D},
    doi = {10.1051/0004-6361/202453539},
    urldate = {2025-11-25},
    journal = {\aap},
    author = {Delfini, L. and Vioque, M. and Ribas, {\'A}. and Hodgkin, S.},
    month = jul,
    year = {2025},
    xnote = {Publisher: EDP ADS Bibcode: 2025A\&A...699A.145D},
    pages = {A145},
}

@article{silverberg2020,
    title = {Peter {Pan} {Disks}: {Long}-lived {Accretion} {Disks} {Around} {Young} {M} {Stars}},
    volume = {890},
    issn = {0004-637X},
    shorttitle = {Peter {Pan} {Disks}},
    url = {https://ui.adsabs.harvard.edu/abs/2020ApJ...890..106S},
    doi = {10.3847/1538-4357/ab68e6},
    urldate = {2025-11-26},
    journal = {\apj},
    author = {Silverberg, Steven M. and Wisniewski, John P. and Kuchner, Marc J. and Lawson, Kellen D. and Bans, Alissa S. and Debes, John H. and Biggs, Joseph R. and Bosch, Milton K. D. and Doll, Katharina and Luca, Hugo A. Durantini and Enachioaie, Alexandru and Hamilton, Joshua and Holden, Jonathan and Hyogo, Michiharu},
    month = feb,
    year = {2020},
    xnote = {Publisher: IOP ADS Bibcode: 2020ApJ...890..106S},
    pages = {106},
}

@ARTICLE{Dullemond2006,
       author = {{Dullemond}, C.~P. and {Natta}, A. and {Testi}, L.},
        title = "{Accretion in Protoplanetary Disks: The Imprint of Core Properties}",
      journal = {\apjl},
         year = 2006,
        month = jul,
       volume = {645},
       number = {1},
        pages = {L69-L72},
          doi = {10.1086/505744},
archivePrefix = {arXiv},
       eprint = {astro-ph/0605336},
 primaryClass = {astro-ph},
       adsurl = {https://ui.adsabs.harvard.edu/abs/2006ApJ...645L..69D}
}

@ARTICLE{Alexander2006,
       author = {{Alexander}, R.~D. and {Clarke}, C.~J. and {Pringle}, J.~E.},
        title = "{Photoevaporation of protoplanetary discs - II. Evolutionary models and observable properties}",
      journal = {\mnras},
         year = 2006,
        month = jun,
       volume = {369},
       number = {1},
        pages = {229-239},
          doi = {10.1111/j.1365-2966.2006.10294.x},
archivePrefix = {arXiv},
       eprint = {astro-ph/0603254},
 primaryClass = {astro-ph},
       adsurl = {https://ui.adsabs.harvard.edu/abs/2006MNRAS.369..229A}
}

@ARTICLE{Villenave2020,
       author = {{Villenave}, M. and {M{\'e}nard}, F. and {Dent}, W.~R.~F. and {Duch{\^e}ne}, G. and {Stapelfeldt}, K.~R. and {Benisty}, M. and {Boehler}, Y. and {van der Plas}, G. and {Pinte}, C. and {Telkamp}, Z. and {Wolff}, S. and {Flores}, C. and {Lesur}, G. and {Louvet}, F. and {Riols}, A. and {Dougados}, C. and {Williams}, H. and {Padgett}, D.},
        title = "{Observations of edge-on protoplanetary disks with ALMA. I. Results from continuum data}",
      journal = {\aap},
         year = 2020,
        month = oct,
       volume = {642},
          eid = {A164},
        pages = {A164},
          doi = {10.1051/0004-6361/202038087},
archivePrefix = {arXiv},
       eprint = {2008.06518},
 primaryClass = {astro-ph.SR},
       adsurl = {https://ui.adsabs.harvard.edu/abs/2020A&A...642A.164V}
}

@ARTICLE{Villenave2022,
       author = {{Villenave}, M. and {Stapelfeldt}, K.~R. and {Duch{\^e}ne}, G. and {M{\'e}nard}, F. and {Lambrechts}, M. and {Sierra}, A. and {Flores}, C. and {Dent}, W.~R.~F. and {Wolff}, S. and {Ribas}, {\'A}. and {Benisty}, M. and {Cuello}, N. and {Pinte}, C.},
        title = "{A Highly Settled Disk around Oph163131}",
      journal = {\apj},
         year = 2022,
        month = may,
       volume = {930},
       number = {1},
          eid = {11},
        pages = {11},
          doi = {10.3847/1538-4357/ac5fae},
archivePrefix = {arXiv},
       eprint = {2204.00640},
 primaryClass = {astro-ph.SR},
       adsurl = {https://ui.adsabs.harvard.edu/abs/2022ApJ...930...11V}
}

@ARTICLE{Haisch2001,
       author = {{Haisch}, Jr., Karl E. and {Lada}, Elizabeth A. and {Lada}, Charles J.},
        title = "{Disk Frequencies and Lifetimes in Young Clusters}",
      journal = {\apjl},
         year = 2001,
        month = jun,
       volume = {553},
       number = {2},
        pages = {L153-L156},
          doi = {10.1086/320685},
archivePrefix = {arXiv},
       eprint = {astro-ph/0104347},
 primaryClass = {astro-ph},
       adsurl = {https://ui.adsabs.harvard.edu/abs/2001ApJ...553L.153H}
}

@ARTICLE{2020A&A.A.R,
       author = {{Ribas}, {\'A}. and {Espaillat}, C.~C. and {Mac{\'\i}as}, E. and {Sarro}, L.~M.},
        title = "{Modeling protoplanetary disk SEDs with artificial neural networks. Revisiting the viscous disk model and updated disk masses}",
      journal = {\aap},
         year = 2020,
        month = oct,
       volume = {642},
          eid = {A171},
        pages = {A171},
          doi = {10.1051/0004-6361/202038352},
archivePrefix = {arXiv},
       eprint = {2009.03323},
 primaryClass = {astro-ph.SR},
       adsurl = {https://ui.adsabs.harvard.edu/abs/2020A&A...642A.171R}
}

@ARTICLE{Picogna2019,
       author = {{Picogna}, Giovanni and {Ercolano}, Barbara and {Owen}, James E. and {Weber}, Michael L.},
        title = "{The dispersal of protoplanetary discs - I. A new generation of X-ray photoevaporation models}",
      journal = {\mnras},
         year = 2019,
        month = jul,
       volume = {487},
       number = {1},
        pages = {691-701},
          doi = {10.1093/mnras/stz1166},
archivePrefix = {arXiv},
       eprint = {1904.02752},
 primaryClass = {astro-ph.EP},
       adsurl = {https://ui.adsabs.harvard.edu/abs/2019MNRAS.487..691P}
}

@ARTICLE{Coleman2024,
       author = {{Coleman}, Gavin A.~L. and {Mroueh}, Joseph K. and {Haworth}, Thomas J.},
        title = "{Photoevaporation obfuscates the distinction between wind and viscous angular momentum transport in protoplanetary discs}",
      journal = {\mnras},
         year = 2024,
        month = jan,
       volume = {527},
       number = {3},
        pages = {7588-7602},
          doi = {10.1093/mnras/stad3692},
archivePrefix = {arXiv},
       eprint = {2311.15824},
 primaryClass = {astro-ph.EP},
       adsurl = {https://ui.adsabs.harvard.edu/abs/2024MNRAS.527.7588C}
}

@ARTICLE{Weder2025,
       author = {{Weder}, Jesse and {Winter}, Andrew J. and {Mordasini}, Christoph},
        title = "{Inferring the physics of protoplanetary disc evolution from the irradiated Cygnus OB2 region: A comparison of viscous and MHD wind-driven scenarios}",
      journal = {\aap},
         year = 2026,
        month = jan,
       volume = {705},
          eid = {A102},
        pages = {A102},
          doi = {10.1051/0004-6361/202556316},
archivePrefix = {arXiv},
       eprint = {2511.01972},
 primaryClass = {astro-ph.EP},
       adsurl = {https://ui.adsabs.harvard.edu/abs/2026A&A...705A.102W}
}

@ARTICLE{Toci2021,
       author = {{Toci}, Claudia and {Rosotti}, Giovanni and {Lodato}, Giuseppe and {Testi}, Leonardo and {Trapman}, Leon},
        title = "{On the secular evolution of the ratio between gas and dust radii in protoplanetary discs}",
      journal = {\mnras},
         year = 2021,
        month = oct,
       volume = {507},
       number = {1},
        pages = {818-833},
          doi = {10.1093/mnras/stab2112},
archivePrefix = {arXiv},
       eprint = {2107.09914},
 primaryClass = {astro-ph.EP},
       adsurl = {https://ui.adsabs.harvard.edu/abs/2021MNRAS.507..818T}
}

@ARTICLE{Delussu2024,
       author = {{Delussu}, Luca and {Birnstiel}, Tilman and {Miotello}, Anna and {Pinilla}, Paola and {Rosotti}, Giovanni and {Andrews}, Sean M.},
        title = "{Population synthesis models indicate a need for early and ubiquitous disk substructures}",
      journal = {\aap},
         year = 2024,
        month = aug,
       volume = {688},
          eid = {A81},
        pages = {A81},
          doi = {10.1051/0004-6361/202450328},
archivePrefix = {arXiv},
       eprint = {2405.14501},
 primaryClass = {astro-ph.EP},
       adsurl = {https://ui.adsabs.harvard.edu/abs/2024A&A...688A..81D}
}

@ARTICLE{Schib2021,
       author = {{Schib}, O. and {Mordasini}, C. and {Wenger}, N. and {Marleau}, G.-D. and {Helled}, R.},
        title = "{The influence of infall on the properties of protoplanetary discs. Statistics of masses, sizes, lifetimes, and fragmentation}",
      journal = {\aap},
         year = 2021,
        month = jan,
       volume = {645},
          eid = {A43},
        pages = {A43},
          doi = {10.1051/0004-6361/202039154},
archivePrefix = {arXiv},
       eprint = {2011.05996},
 primaryClass = {astro-ph.EP},
       adsurl = {https://ui.adsabs.harvard.edu/abs/2021A&A...645A..43S}
}

@ARTICLE{Schib2025,
       author = {{Schib}, O. and {Mordasini}, C. and {Emsenhuber}, A. and {Helled}, R.},
        title = "{DIPSY: A new Disc Instability Population SYnthesis: I. Modelling, evolution of individual systems, and tests}",
      journal = {\aap},
         year = 2025,
        month = dec,
       volume = {704},
          eid = {A27},
        pages = {A27},
          doi = {10.1051/0004-6361/202556262},
archivePrefix = {arXiv},
       eprint = {2510.02436},
 primaryClass = {astro-ph.EP},
       adsurl = {https://ui.adsabs.harvard.edu/abs/2025A&A...704A..27S}
}

@ARTICLE{Zagaria2022,
       author = {{Zagaria}, Francesco and {Clarke}, Cathie J. and {Rosotti}, Giovanni P. and {Manara}, Carlo F.},
        title = "{Stellar multiplicity affects the correlation between protoplanetary disc masses and accretion rates: binaries explain high accretors in Upper Sco}",
      journal = {\mnras},
         year = 2022,
        month = may,
       volume = {512},
       number = {3},
        pages = {3538-3550},
          doi = {10.1093/mnras/stac621},
archivePrefix = {arXiv},
       eprint = {2203.01986},
 primaryClass = {astro-ph.SR},
       adsurl = {https://ui.adsabs.harvard.edu/abs/2022MNRAS.512.3538Z}
}

@ARTICLE{Downes2015,
       author = {{Downes}, Juan Jos{\'e} and {Rom{\'a}n-Z{\'u}{\~n}iga}, Carlos and {Ballesteros-Paredes}, Javier and {Mateu}, Cecilia and {Brice{\~n}o}, C{\'e}sar and {Hern{\'a}ndez}, Jes{\'u}s and {Petr-Gotzens}, Monika G. and {Calvet}, Nuria and {Hartmann}, Lee and {Mauco}, Karina},
        title = "{The number fraction of discs around brown dwarfs in Orion OB1a and the 25 Orionis group}",
      journal = {\mnras},
         year = 2015,
        month = jul,
       volume = {450},
       number = {4},
        pages = {3490-3502},
          doi = {10.1093/mnras/stv888},
       adsurl = {https://ui.adsabs.harvard.edu/abs/2015MNRAS.450.3490D}
}

@ARTICLE{Ratzenbock2023,
       author = {{Ratzenb{\"o}ck}, Sebastian and {Gro{\ss}schedl}, Josefa E. and {Alves}, Jo{\~a}o and {Miret-Roig}, N{\'u}ria and {Bomze}, Immanuel and {Forbes}, John and {Goodman}, Alyssa and {Hacar}, {\'A}lvaro and {Lin}, Doug and {Meingast}, Stefan and {M{\"o}ller}, Torsten and {Piecka}, Martin and {Posch}, Laura and {Rottensteiner}, Alena and {Swiggum}, Cameren and {Zucker}, Catherine},
        title = "{The star formation history of the Sco-Cen association. Coherent star formation patterns in space and time}",
      journal = {\aap},
         year = 2023,
        month = oct,
       volume = {678},
          eid = {A71},
        pages = {A71},
          doi = {10.1051/0004-6361/202346901},
archivePrefix = {arXiv},
       eprint = {2302.07853},
 primaryClass = {astro-ph.SR},
       adsurl = {https://ui.adsabs.harvard.edu/abs/2023A&A...678A..71R}
}

@ARTICLE{Winter2018,
       author = {{Winter}, A.~J. and {Clarke}, C.~J. and {Rosotti}, G. and {Ih}, J. and {Facchini}, S. and {Haworth}, T.~J.},
        title = "{Protoplanetary disc truncation mechanisms in stellar clusters: comparing external photoevaporation and tidal encounters}",
      journal = {\mnras},
         year = 2018,
        month = aug,
       volume = {478},
       number = {2},
        pages = {2700-2722},
          doi = {10.1093/mnras/sty984},
archivePrefix = {arXiv},
       eprint = {1804.00013},
 primaryClass = {astro-ph.SR},
       adsurl = {https://ui.adsabs.harvard.edu/abs/2018MNRAS.478.2700W}
}

@ARTICLE{Eisner2018,
       author = {{Eisner}, J.~A. and {Arce}, H.~G. and {Ballering}, N.~P. and {Bally}, J. and {Andrews}, S.~M. and {Boyden}, R.~D. and {Di Francesco}, J. and {Fang}, M. and {Johnstone}, D. and {Kim}, J.~S. and {Mann}, R.~K. and {Matthews}, B. and {Pascucci}, I. and {Ricci}, L. and {Sheehan}, P.~D. and {Williams}, J.~P.},
        title = "{Protoplanetary Disk Properties in the Orion Nebula Cluster: Initial Results from Deep, High-resolution ALMA Observations}",
      journal = {\apj},
         year = 2018,
        month = jun,
       volume = {860},
       number = {1},
          eid = {77},
        pages = {77},
          doi = {10.3847/1538-4357/aac3e2},
archivePrefix = {arXiv},
       eprint = {1805.03669},
 primaryClass = {astro-ph.SR},
       adsurl = {https://ui.adsabs.harvard.edu/abs/2018ApJ...860...77E}
}

\begin{appendix} 
\section{Initial distributions}
\label{app:initial_dist}

In this appendix, we present the initial distributions for the star formation rate and the gas disc masses adopted in our population synthesis models.

\begin{figure}[h]
\centering
\includegraphics[width=\hsize]{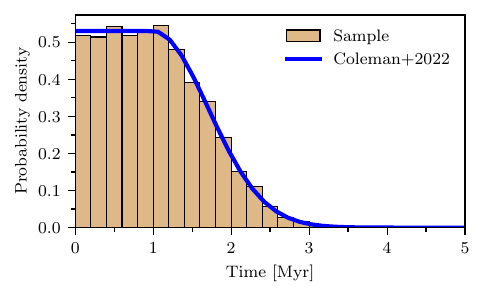}
  \caption{The blue line represents the probability density of the star formation rate from \citet{coleman2022dispersal}. The light brown histograms show the discrete probability applied to our population synthesis.}
    \label{Fig:SFR}
\end{figure}

\begin{figure}[h]
    \centering
    \includegraphics[width=\hsize]{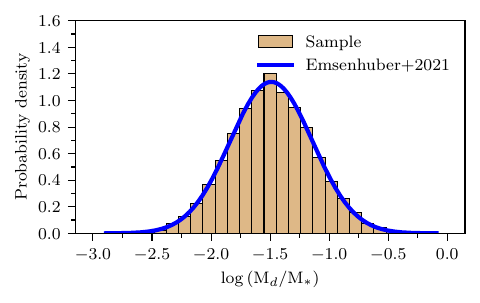}
    \caption{The blue line represents the probability density of the gas disc masses (in terms of the mass of the host star) obtained by \citet{tychoniec2018vla}. The light brown histograms correspond to the disc mass sample for our population synthesis.}
    \label{Fig:DensFuntMd}
\end{figure}

\section{Photoevaporation models}
\label{app:photoevaporation}

In this appendix, we provide the explicit equations used to model the mass-loss rates due to internal and external photoevaporation, which are incorporated into the diffusion equation (Eq.~\ref{dens_sup_completa}) through the sink term $\dot{\Sigma}_{\rm w} = \dot{\Sigma}_{\rm XR} + \dot{\Sigma}_{\rm FUV} + \dot{\Sigma}_{\rm ext~fuv}$.

\subsection{Internal photoevaporation}

The mass loss due to X-ray (XR) photoevaporation from the central star is computed following the analytical prescriptions by \cite{owen2012theoryfoto}:
\begin{equation}
    \dot{M}_{\rm XR}=  c_0\times 10 ^{-9}\left(\frac{M_{\star}}{1M_{\odot}}\right)^{c_1} \times 
    \left(\frac{L_{\rm XR}}{10^{30} \rm{erg}~\rm{s}^{-1}} \right)^{1.14}\rm{M}_{\odot} \rm{ ~yr}^{-1},
    \label{eq:Mpunto_app}
\end{equation}    
where $c_0=6.25$ and $c_1=-0.068$ for the primordial disc stage. When a hole is fully opened in the inner part of the disc, $c_0=4.8$ and $c_1=-0.148$. The XR luminosity is given by:
\begin{equation}
    \log(L_{\rm XR}[\rm erg\,s^{-1}]) = 30.37 + 1.44\log\left( \frac{{M}_{\star}}{{M}_{\odot}}\right).
    \label{eq:lum_RX_app}
\end{equation}

The rate of far ultraviolet (FUV) photoevaporation from the central star is computed following \citet{Kunitomo2021}, where:
\begin{align}
    \dot{\Sigma}_{\rm{FUV}}(R\geq R _{\rm{FUV}}) = \dot{\Sigma}_{0} \left( \frac{L _{\rm{FUV}}}{10^{31.7}\rm{erg}~\rm{s}^{-1}} \right) \left( \frac{R}{4\rm{~au}} \right)^{-2},
\end{align}
and $\dot{ \Sigma }_{ \rm{FUV} }(R<{R}_{\rm{FUV}}) = 0$. Here, $R_{\rm{FUV}} = 4 ({M}_{\star}/ {M}_{\odot})$~au, and $\dot{\Sigma}_{0} = 10^{-12}~\rm{g}~\rm{cm}^{-2}\rm{s}^{-1}$. The term $L_{\rm{FUV}}$ includes contributions from stellar accretion, the photosphere, and the chromosphere.

\subsection{External photoevaporation}

Following \cite{coleman2022dispersal}, disc mass is removed from the outer regions due to external FUV irradiation. The corresponding mass-loss rate profile is given by:
\begin{align}
    \dot{\Sigma}_{\rm{ext ~ fuv}} = G_{\rm sm} \frac{ \dot{M}_{\rm d}^{\rm{fuv} } (R_{\rm d})}{ \pi \left(R_{\rm d}^2 - \beta^2 R_{\rm{g,~fuv} }^2\right)},
    \label{eq:sigmapexter_app}
\end{align}
where $\beta =0.14$ and $R_{\rm d}$ is the disc distance where $\Sigma_{\rm g}(R)>10^{-4} \rm g \, \rm cm^{-2}$. The factor $G_{\rm sm}$ is a smoothing function around the effective gravitational radius $R_{\rm g,~fuv}$:
\begin{align}
    G_{\rm sm} = 1 - \left( 1+ \left( \frac{R_{\rm d}}{\beta R_{\rm g, ~fuv}}\right)^{20} \right)^{-1}.
\end{align}
The total external mass-loss rate $\dot{M}^{\rm fuv}_{\rm d}$ scales linearly with the outer radius of the disc and is parametrized by:
\begin{equation}
    \dot{M}^{\rm fuv}_{\rm d} = \dot{M}_{\rm d}^{fuv,0} \frac{R_{\rm d}}{100 \, \text{au}},
\end{equation}
where $\dot{M}_{\rm d}^{fuv,0}$ is treated as a free parameter associated with the external radiation environment.

\section{Quantitative fit results}
\label{app:tables} 

In this appendix, we present the comprehensive quantitative metrics used to evaluate the performance of our population synthesis models against observational data. Table \ref{tab:regresiones_pendientes_mstar} and \ref{tab:regresiones_pendientes_md}  details the linear regression parameters (slope and intercept) for the mass accretion rate correlations, comparing our synthetic results directly with the empirical fits derived by \cite{manara2023ASPC}. Furthermore, Table \ref{tab:lifetimes} summarizes the mean disc lifetimes ($\tau$) obtained for each tested model, allowing for a quantitative comparison with the observational constraints provided by \cite{Pfalzner_2022}.

\begin{table*}[ht!]
    \centering
    \caption{Linear regression coefficients for the accretion rate correlation with stellar mass ($\log \dot{M}$ vs. $\log M_\star$).}
    \label{tab:regresiones_pendientes_mstar}
    \renewcommand{\arraystretch}{1.2} 
    \begin{tabular}{l c c c c}
        \toprule
        \textbf{synthesis} & \textbf{ $log10$ $a_0$} & \textbf{Error $a_0$} & \textbf{$a_1$ } & \textbf{Error $a_1$} \\
        \hline
        \multirow{1}{*}{Observational data \cite{manara2023ASPC}} 
        & -7.96  & 0.05 & $1.8231$ & $0.0072$ \\
        \hline
        \multirow{1}{*}{Reference Without SFR } 
        & -9.24 & 0.01 & 0.3052 & 0.0005 \\
        \multirow{1}{*}{ Reference With SFR } 
        & -8.81 & 0.01 & 0.3150 & 0.0003 \\
        \hline
        \multirow{1}{*}{ $\alpha = \alpha_{0} \,  M_{\star}^{1.8}$, log $\alpha_{0}$ U(-4,-2)} 
        & -8.995 & 0.009 & 1.5496 & 0.0003 \\
        \multirow{1}{*}{ $\alpha = \alpha_{0} \,  M_{\star}^{1.8}$, log $\alpha_{0}$ linear(-4,-2)} 
        & -8.585 & 0.008 &1.7382 & 0.0003 \\
        \hline
        \multirow{1}{*}{ $\alpha = \alpha_{0} \,  M_{\star}^{1.8}$, log $\alpha_{0}$ U(-3.5,-1.5)} 
        & -8.428 & 0.01 & 1.7326 & 0.0003 \\
        \multirow{1}{*}{ $\alpha = \alpha_{0} \,  M_{\star}^{1.8}$, log $\alpha_{0}$ linear(-3.5,-1.5)} 
        & -7.99 & 0.01 & 1.8751 & 0.0003 \\
        \hline
        \multirow{1}{*}{$\dot{M}_{d}^{fuv,0} = 10^{-9} M_{\odot}yr^{-1}  $}
        & -7.985 & 0.009 & 1.8803 & 0.0003 \\
        \multirow{1}{*}{$\dot{M}_{d}^{fuv,0} = 10^{-8} M_{\odot}yr^{-1}  $} 
        & -8.04 & 0.01 & 1.8793 & 0.0004 \\
        \multirow{1}{*}{$\dot{M}_{d}^{fuv,0} = 10^{-7} M_{\odot}yr^{-1}  $} 
        & -8.23 & 0.01 & 1.824 & 0.0008 \\
        \hline
        \multirow{1}{*}{ $\alpha = \alpha_{0} \,  M_{\star}^{1.5}$, log $\alpha_{0}$ U(-4,-2)} 
        & -8.9658 & 0.009 & 1.3613 & 0.0003 \\
        \multirow{1}{*}{ $\alpha = \alpha_{0} \,  M_{\star}^{1.5}$, log $\alpha_{0}$ linear(-4,-2)} 
        & -8.554 & 0.009 & 1.4924 & 0.0003 \\
        \hline
        \multirow{1}{*}{ $\alpha = \alpha_{0} \,  M_{\star}^{2.0}$, log $\alpha_{0}$ U(-4,-2)} 
        & -9.003 & 0.009 &  1.6852 & 0.0003 \\
        \multirow{1}{*}{ $\alpha = \alpha_{0} \,  M_{\star}^{2.0}$, log $\alpha_{0}$ linear(-4,-2)} 
        & -8.587 & 0.008 & 1.8995 & 0.0003 \\
        \hline
        \multirow{1}{*}{ $\alpha = \alpha_{0} \,  M_{\star}^{1.5}$, log $\alpha_{0}$ U(-3.5,-1.5)} 
        & -8.38 & 0.01 & 1.5235 & 0.0003 \\
        \multirow{1}{*}{ $\alpha = \alpha_{0} \,  M_{\star}^{1.5}$, log $\alpha_{0}$ linear(-3.5,-1.5)} 
        & -7.959 & 0.009 & 1.6530 & 0.0003 \\
        \hline
        \multirow{1}{*}{ $\alpha = \alpha_{0} \,  M_{\star}^{2.0}$, log $\alpha_{0}$ U(-3.5,-1.5)} 
        & -8.406 & 0.009 & 1.9049 & 0.0003 \\
        \multirow{1}{*}{ $\alpha = \alpha_{0} \,  M_{\star}^{2.0}$, log $\alpha_{0}$ linear(-3.5,-1.5)} 
        & -8.01 & 0.01 & 2.0242 & 0.0003 \\
        \hline
    \end{tabular}
    \tablefoot{The table lists the intercept $a_0$ and slope $a_1$ alongside their standard errors. The table contrasts the empirical constraints from \cite{manara2023ASPC} with the outcomes of our various synthetic population models under different physical assumptions.}
    \tablefoot{}
\end{table*}

\begin{table*}[ht!]
    \centering
    \caption{ Linear regression coefficients for the accretion rate correlation with disc mass ($\log \dot{M}$ vs. $\log M_d$).}
    \label{tab:regresiones_pendientes_md}
    \renewcommand{\arraystretch}{1.2} 
    \begin{tabular}{l c c c c}
        \toprule
        \textbf{synthesis} & \textbf{ $log10$ $a_0$} & \textbf{Error $a_0$} & \textbf{$a_1$ } & \textbf{Error $a_1$} \\
        \hline
        \multirow{1}{*}{Observational data \cite{manara2023ASPC}} 
        & -6.20 & 0.05 & 0.895 & 0.003 \\
        \hline
        \multirow{1}{*}{Reference Without SFR } 
        & -9.24 & 0.01 & 0.3052 & 0.0005 \\
        \multirow{1}{*}{ Reference With SFR } 
        & -7.86 & 0.01 & 0.5025 & 0.0002  \\
        \hline
        \multirow{1}{*}{ $\alpha = \alpha_{0} \,  M_{\star}^{1.8}$, log $\alpha_{0}$ U(-4,-2)} 
        & -7.645 & 0.009 & 1.0966 & 0.0002 \\
        \multirow{1}{*}{ $\alpha = \alpha_{0} \,  M_{\star}^{1.8}$, log $\alpha_{0}$ linear(-4,-2)} 
        & -7.094 & 0.008 & 1.2289 & 0.0002 \\
        \hline
        \multirow{1}{*}{ $\alpha = \alpha_{0} \,  M_{\star}^{1.8}$, log $\alpha_{0}$ U(-3.5,-1.5)} 
        & -6.947 & 0.01 & 1.2157 & 0.0003 \\
        \multirow{1}{*}{ $\alpha = \alpha_{0} \,  M_{\star}^{1.8}$, log $\alpha_{0}$ linear(-3.5,-1.5)} 
        & -6.447 & 0.009 & 1.2946 & 0.0002 \\
        \hline
        \multirow{1}{*}{$\dot{M}_{d}^{fuv,0} = 10^{-9} M_{\odot}yr^{-1}  $}
        & -6.491 & 0.009 & 1.2393 & 0.0002 \\
         \multirow{1}{*}{$\dot{M}_{d}^{fuv,0} = 10^{-8} M_{\odot}yr^{-1}  $} 
        & -7.46 & 0.01 & 0.6556 & 0.0002 \\
         \multirow{1}{*}{$\dot{M}_{d}^{fuv,0} = 10^{-7} M_{\odot}yr^{-1}  $} 
        & -7.19 & 0.01 & 0.6471 & 0.0002 \\
        \hline
        \multirow{1}{*}{Data \cite{manara2023ASPC}: 25\%} 
        & -6.74 & 0.05 & 0.895 & 0.004 \\
        \multirow{1}{*}{Data \cite{manara2023ASPC}: 10\%} 
        & -7.09 & 0.05 & 0.895 & 0.004 \\
        \hline
        
        \multirow{1}{*}{ $\alpha = \alpha_{0} \,  M_{\star}^{1.5}$, log $\alpha_{0}$ U(-4,-2)} 
        & -7.667 & 0.009 & 1.0155 & 0.0002 \\
        \multirow{1}{*}{ $\alpha = \alpha_{0} \,  M_{\star}^{1.5}$, log $\alpha_{0}$ linear(-4,-2)} 
        & -7.098 & 0.008 & 1.1292 & 0.0002 \\
        \hline
        
        \multirow{1}{*}{ $\alpha = \alpha_{0} \,  M_{\star}^{2.0}$, log $\alpha_{0}$ U(-4,-2)} 
        & -7.666 & 0.009 & 1.1328 & 0.0002 \\
        \multirow{1}{*}{ $\alpha = \alpha_{0} \,  M_{\star}^{2.0}$, log $\alpha_{0}$ linear(-4,-2)} 
        & -7.075 & 0.009 & 1.2856 & 0.0002 \\
        \hline
        \multirow{1}{*}{ $\alpha = \alpha_{0} \,  M_{\star}^{1.5}$, log $\alpha_{0}$ U(-3.5,-1.5)} 
        & -7.01 & 0.01 & 1.1012 & 0.0002 \\
        \multirow{1}{*}{ $\alpha = \alpha_{0} \,  M_{\star}^{1.5}$, log $\alpha_{0}$ linear(-3.5,-1.5)} 
        & -6.5 & 0.01 & 1.1752 & 0.0002 \\
        \hline
        \multirow{1}{*}{ $\alpha = \alpha_{0} \,  M_{\star}^{2.0}$, log $\alpha_{0}$ U(-3.5,-1.5)} 
        & -7.02 & 0.01 & 1.2381 & 0.0003 \\
        \multirow{1}{*}{ $\alpha = \alpha_{0} \,  M_{\star}^{2.0}$, log $\alpha_{0}$ linear(-3.5,-1.5)} 
        & -6.41 & 0.01 & 1.3749 & 0.0003 \\
        \hline
    \end{tabular}
    \tablefoot{Similar to Table.\ref{tab:regresiones_pendientes_mstar}.}
\end{table*}

\begin{table*}[ht!]
    \centering
    \caption{Mean disc lifetimes ($\tau$) and their associated standard errors derived from the synthetic populations. }
    \label{tab:lifetimes}
    \renewcommand{\arraystretch}{1.2} 
    \begin{tabular}{l c c c }
        \toprule
        \textbf{synthesis} & $\tau$ & \textbf{Error } \\
        \hline
        \multirow{1}{*}{Reference value \citep{Pfalzner_2022}} 
        & 7 & --- \\
        \hline
        \multirow{1}{*}{Reference Without SFR } 
        & 3.99 & 0.01 \\
        \multirow{1}{*}{Reference With SFR } 
        & 5.08  & 0.01 \\
        \hline
        \multirow{1}{*}{ $\alpha = \alpha_{0} \,  M_{\star}^{1.8}$, log $\alpha_{0}$ U(-4,-2)} 
        & 8.16 & 0.02 \\
        \multirow{1}{*}{ $\alpha = \alpha_{0} \,  M_{\star}^{1.8}$, log $\alpha_{0}$ linear(-4,-2)} 
        & 7.79 & 0.01  \\
        \hline
        \multirow{1}{*}{ $\alpha = \alpha_{0} \,  M_{\star}^{1.8}$, log $\alpha_{0}$ U(-3.5,-1.5)} 
        & 7.15  & 0.01  \\
        \multirow{1}{*}{ $\alpha = \alpha_{0} \,  M_{\star}^{1.8}$, log $\alpha_{0}$ linear(-3.5,-1.5)} 
         & 6.57  & 0.01  \\
        \hline
        \multirow{1}{*}{$\dot{M}_{d}^{fuv,0} = 10^{-9} M_{\odot}yr^{-1}  $} 
         & 4.12 & 0.02  \\
         \multirow{1}{*}{$\dot{M}_{d}^{fuv,0} = 10^{-8} M_{\odot}yr^{-1}  $} 
         & 1.92 & 0.01  \\
         \multirow{1}{*}{$\dot{M}_{d}^{fuv,0} = 10^{-7} M_{\odot}yr^{-1}  $} 
         & 1.389 & 0.005  \\
        \hline
        \multirow{1}{*}{ Upper Scorpius case } 
         & 2.691 & 0.004 \\
        \hline
        \multirow{1}{*}{ $\alpha = \alpha_{0} \,  M_{\star}^{1.5}$, log $\alpha_{0}$ U(-4,-2)} 
         & 7.80 & 0.02 \\
        \multirow{1}{*}{ $\alpha = \alpha_{0} \,  M_{\star}^{1.5}$, log $\alpha_{0}$ linear(-4,-2)} 
         & 7.29 & 0.01 \\
        \hline
        \multirow{1}{*}{ $\alpha = \alpha_{0} \,  M_{\star}^{2.0}$, log $\alpha_{0}$ U(-4,-2)} 
         & 8.40 & 0.02 \\
        \multirow{1}{*}{ $\alpha = \alpha_{0} \,  M_{\star}^{2.0}$, log $\alpha_{0}$ linear(-4,-2)} 
         & 7.84 & 0.01  \\
        \hline
        \multirow{1}{*}{ $\alpha = \alpha_{0} \,  M_{\star}^{1.5}$, log $\alpha_{0}$ U(-3.5,-1.5)} 
        & 6.71 & 0.01 \\
        \multirow{1}{*}{ $\alpha = \alpha_{0} \,  M_{\star}^{1.5}$, log $\alpha_{0}$ linear(-3.5,-1.5)} 
        & 5.89 & 0.01 \\
        \hline
        \multirow{1}{*}{ $\alpha = \alpha_{0} \,  M_{\star}^{2.0}$, log $\alpha_{0}$ U(-3.5,-1.5)} 
        & 7.55 & 0.01 \\
        \multirow{1}{*}{ $\alpha = \alpha_{0} \,  M_{\star}^{2.0}$, log $\alpha_{0}$ linear(-3.5,-1.5)} 
        &  6.98 & 0.01 \\
        \hline
    \end{tabular}
    \tablefoot{The empirical reference value from \cite{Pfalzner_2022} is provided at the top for direct comparison with our model outputs.}
\end{table*}

\section{Dependence of the viscosity with the stellar mass}
\label{sec:appendix_viscosity} 

In order to study the impact of the parameter $a$ in the relation $\alpha = \alpha_{0} \times M_{\star}^{a}$, we analyse the results of new synthetic populations adopting now $a =1.5$ and $a=2$. We show the cases considering the previously studied two ranges for $\alpha_0$ and adopting the uniform and linear probability distribution in such ranges as in Sec.~\ref{sec:res-corr1} and Sec. \ref{sec:-3.5-1.5}.

\subsection{$\log \alpha_{0}$ between -4 to -2}
\label{Ap:-4.0-2.0}

\begin{figure*}[!!ht]
    \centering
    \includegraphics[width=0.9\textwidth]{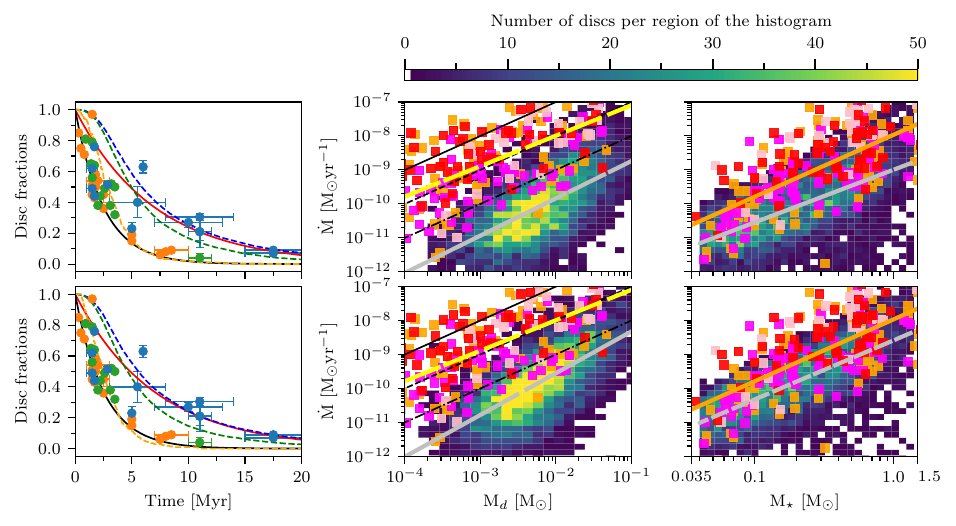}
    \caption{Similar to Fig.\ref{Fig:reference}, but considering the $\alpha$-viscosity parameter as a function of the stellar mass ($\alpha = \alpha_{0} \,  M_{\star}^{1.8}$) and considering an SFR with $t_{\rm d}= 1$~Myr. The top row corresponds to the results for the population synthesis adopting a uniform distribution for $\alpha_0$, while the bottom row corresponds to consider a linear distribution for $\alpha_0$. In both cases the range of $\log \alpha_0$ is between -4 and -2.}
    \label{fig:LR_app}
\end{figure*}

\begin{figure*}[!!ht]
    \centering
\includegraphics[width=0.9\textwidth]{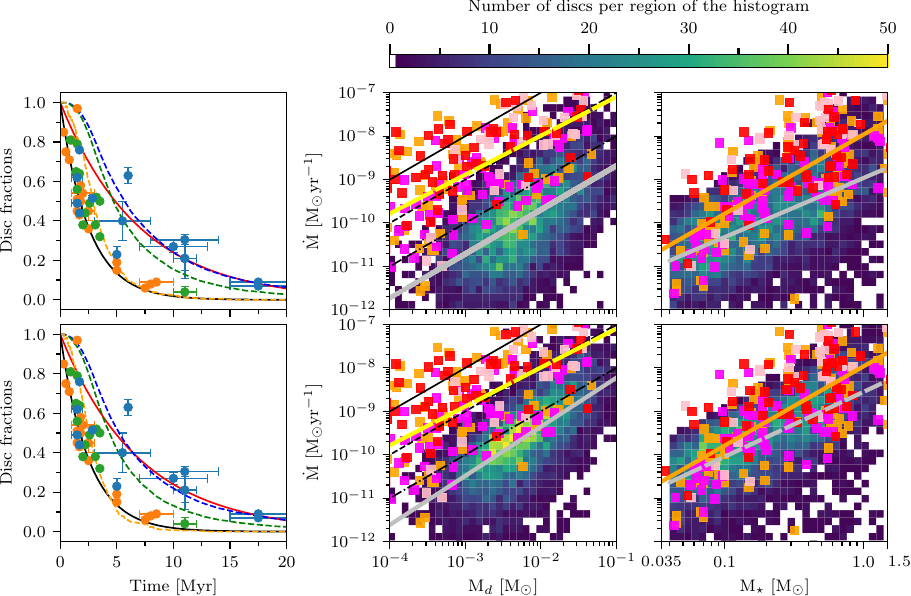}
    \caption{Same as Fig.~\ref{fig:LR_app} but adopting a value of $a=1.5$ in the relation $\alpha = \alpha_{0} \times M_{\star}^{a}$.}
    \label{fig:1.5LR}
\end{figure*}

\begin{figure*}[!!ht]
    \centering
    \includegraphics[width=0.9\linewidth]{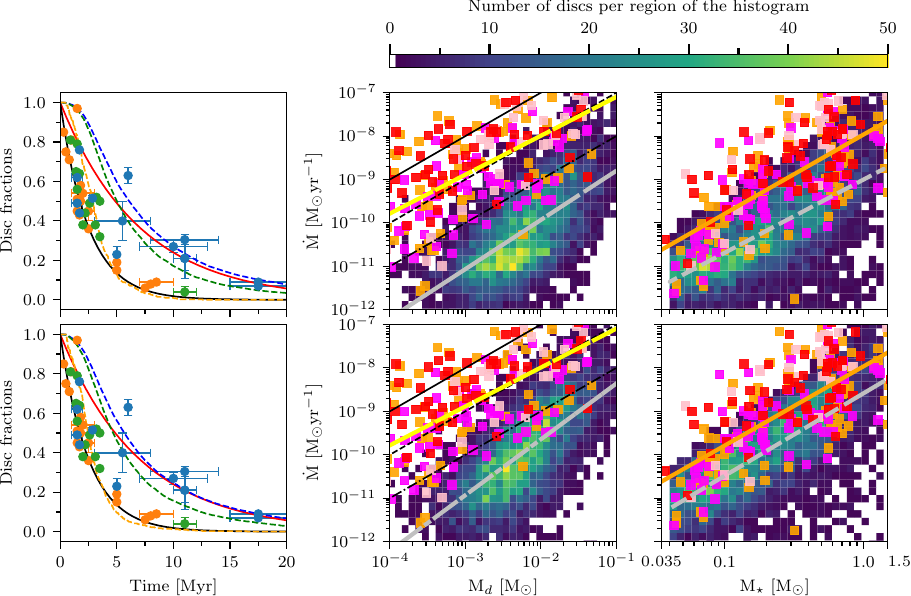}
    \caption{Same as Fig.~\ref{fig:LR_app} but adopting a value of $a= 2$ in the relation $\alpha = \alpha_{0} \times M_{\star}^{a}$.}
    \label{fig:2.0LR}
\end{figure*}

In Figure \ref{fig:LR_app}, we present the results for $a= 1.8$. Furthermore, Figures \ref{fig:1.5LR} and \ref{fig:2.0LR} show the results for $a= 1.5$ and $a= 2$, for the uniform and linear distributions of values of $\alpha_0$, respectively. In both cases, as for the case of $a= 1.8$, the fraction of stars with discs matches very well the estimations from \citet{Pfalzner_2022} for the closer clusters, especially after about 5~Myr, covering disc lifetimes in the range $1 - 20$~Myr, in agreement with observations. Increasing the value of $a$ generates a small reduction in the number of stars with discs at early ages. This is because for $a= 2$ stars more massive than $1~\rm M_{\odot}$ have discs with larger viscosities and evolve faster, being the first to dissipate. While for $a= 2$ stars with masses below $1~\rm M_{\odot}$ have discs with less viscosity, dissipation timescales become larger being dominated by the photoevaporation. Thus, both population syntheses seem very similar after about 5~Myr. 

We can also see that the increment of the exponent $a$ leads to a better correlation between accretion rates and disc and stellar masses. 

Summarizing, we can observe that the distribution of $\log \alpha_{0}$ between -4 and -2 leads to reasonable agreement with the observation for the three values of the exponent $a$. This supports an observational necessity that the disc viscosity scales with the stellar mass. 

\subsection{$\log \alpha_{0}$ between -3.5 to -1.5}
\label{Ape:-3.5-1.5}
 
\begin{figure*}[!!ht]
    \centering
    \includegraphics[width=0.9\textwidth]{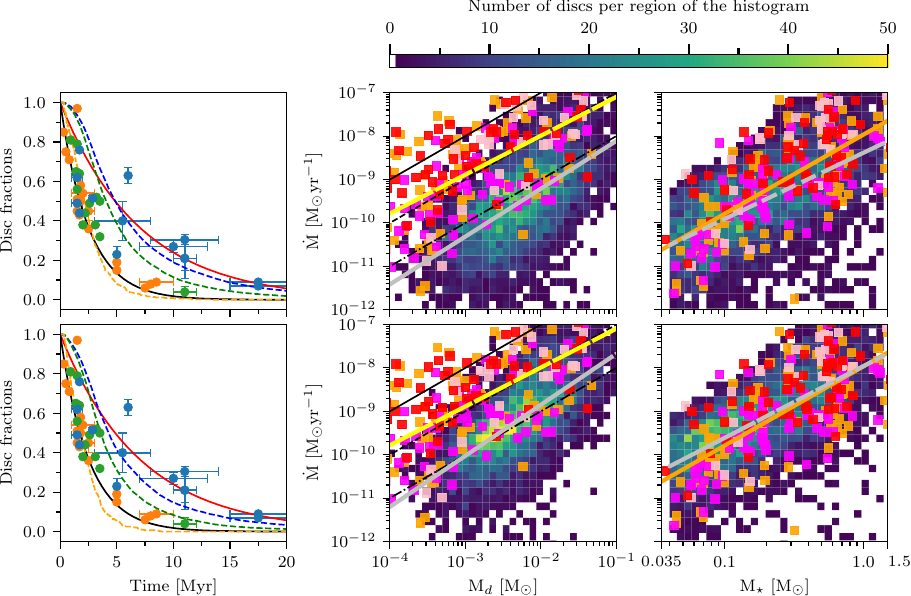}
    \caption{Same as Fig.~\ref{fig:hi} but adopting a value of $a= 1.5$ in the relation $\alpha = \alpha_{0} \times M_{\star}^{a}$.}
    \label{fig:1.5HR}
\end{figure*}

\begin{figure*}[!!ht]
    \centering
    \includegraphics[width=0.9\textwidth]{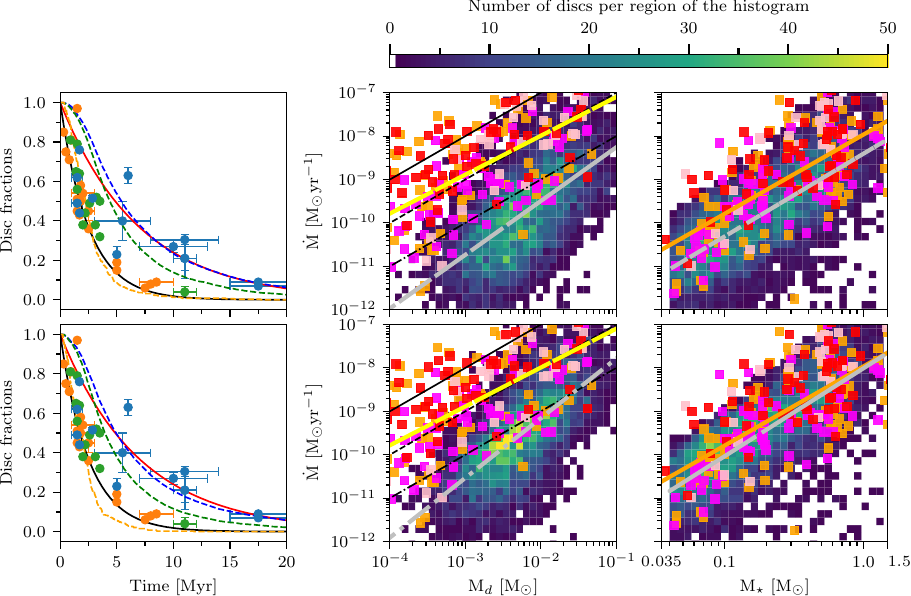}
    \caption{Same as Fig.~\ref{fig:hi} but adopting a value of $a= 2$ in the relation $\alpha = \alpha_{0} \times M_{\star}^{a}$.}
    \label{fig:2.0HR}
\end{figure*}

Finally, we examine the population synthesis assuming a larger viscosity for the discs, in the range of $\log \alpha_{0}$ between -3.5 and -1.5. Again, we consider both the uniform and linear distributions over the range of $\log \alpha_{0}$. Figures \ref{fig:1.5HR} and \ref{fig:2.0HR} show the result for the exponent $a =1.5$ and $a=2.0$, respectively. As before, top row displays the population synthesis considering the uniform distribution of $\log \alpha_{0}$, while the bottom panels represent the case of the linear distribution. 

As we showed in Sec. \ref{sec:-3.5-1.5}, increasing the disc viscosities matches the mass accretion rates as a function of the disc masses better. In addition, it also matches the correlation between the mass accretion rates and the stellar masses for both values of the exponent $a$ and for both distributions of $\log \alpha_{0}$, especially the linear one. 

In this sense, the population synthesis with linear distribution of $\log \alpha_{0}$ and the value of $a= 2$ is the one that better reproduces the fraction of stars with discs. 

As we mentioned before, in the absence of external photoevaporation none of the population syntheses is able to reproduce low-mass discs with high accretion rates. 

Again, the change of the correlation between disc viscosity and stellar mass seems to not play a major role, but it is necessary to reproduce the observational trend between mass accretion rates and stellar masses. 

\section{Time evolution of the correlation between accretion rates and stellar masses}
\label{sec:appendix_time_evo}

The correlation between accretion rates and stellar masses seems to represent a strong observational trend. In this appendix, we want to analyse the time evolution of our synthetic correlation. Thus, for the population synthesis described in Sec.~\ref{sec:-3.5-1.5} (the one adopting a linear distribution for $\log\alpha_0$ between -3.5 and -1.5) we compute the $\dot{M}-M_{\star}$ correlation at different times of the synthesis evolution (in the Sec.~\ref{sec:-3.5-1.5} such correlation was computed at 1.5 Myr of the synthesis evolution). Thus, the panels from left to right represent the synthetic correlation $\dot{M}-M_{\star}$ at 1~Myr, 2~Myr, 3~Myr and 4~Myr, respectively, of the synthesis evolution. It is clear that as time advances, more discs dissipate. However, the synthetic correlation $\dot{M}-M_{\star}$ seems to keep a good match with the observed correlation even at 4~Myr of evolution. This result agrees with the observed correlation between accretion rates and stellar masses being invariant with time \citep[e.g.][]{delfini2025}. In addition, this result also supports the choice to compute synthetic accretion rates at 1.5~Myr as a representative time for young stellar clusters. 

\begin{figure*}[!!ht]
    \centering     
    \includegraphics[width=\textwidth]{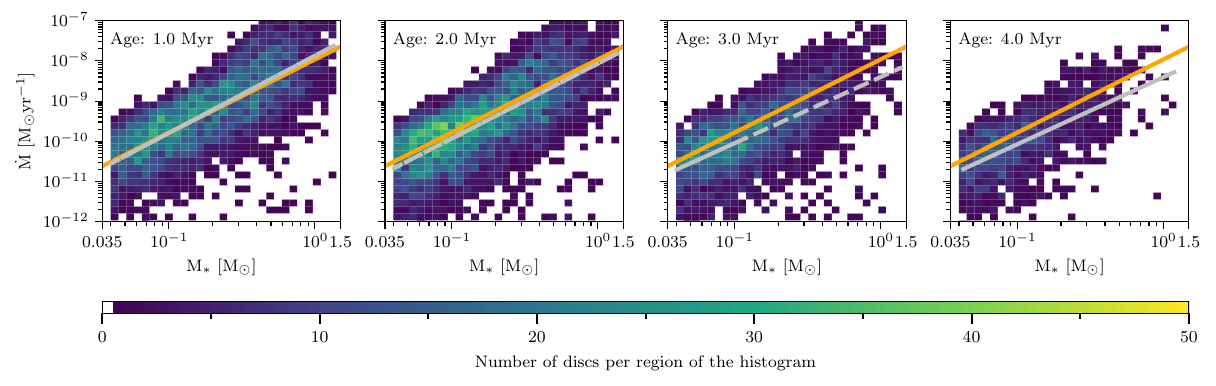}
    \caption{Time evolution of the diagram $\dot{M}-M_{\star}$ for the  population synthesis used in Sec.~\ref{sec:-3.5-1.5}, i.e. considering $a=1.8$, $\log\alpha_0$ following a linear distribution in the range between -3.5 and -1.5, and an SFR with $t_{\rm d}= 1$~Myr.} 
    \label{fig:invtimecorr}
\end{figure*}

\end{appendix}
\end{document}